\documentclass[pra,twocolumn,amssymb,amsmath,unsortedaddress,superscriptaddress]{revtex4-1}
\usepackage{graphicx}
\usepackage{epsfig,color}
\usepackage{amsmath,amssymb,eufrak}
\usepackage{amsthm}
\usepackage{enumerate}
\usepackage{hyperref}

\newcommand{\be}{\begin{equation}}
\newcommand{\ee}{\end{equation}}

\definecolor{cyan}{rgb}{0,0.5,1}
\definecolor{orange}{rgb}{1,0.5,0}

\newcommand{\app}[1]{App.~\ref{#1}}
\newcommand{\eq}[1]{Eq.~\eqref{#1}}
\newcommand{\sect}[1]{Sec.~\ref{#1}}

\newcommand{\ket}[2][]{\left| #2 \right\rangle_{#1}} % use as \ket[ subscript ]{ vector }
\newcommand{\minus}{\! - \!}
\newcommand{\plus}{\! + \!}
\newcommand{\ZZ}{\mathbb{Z}}

\begin{document}

\title{Digital lattice gauge theories}

\date{\today}

\author{Erez Zohar}
\address{Max-Planck-Institut f\"ur Quantenoptik, Hans-Kopfermann-Stra\ss e 1, 85748 Garching, Germany.}

\author{Alessandro Farace}
\address{Max-Planck-Institut f\"ur Quantenoptik, Hans-Kopfermann-Stra\ss e 1, 85748 Garching, Germany.}

\author{Benni Reznik}
\address{School of Physics and Astronomy, Raymond and Beverly Sackler
Faculty of Exact Sciences, Tel Aviv University, Tel-Aviv 69978, Israel.}

\author{J. Ignacio Cirac}
\address{Max-Planck-Institut f\"ur Quantenoptik, Hans-Kopfermann-Stra\ss e 1, 85748 Garching, Germany.}

\begin{abstract}
We propose a general scheme for a digital construction of lattice gauge theories with dynamical fermions. In this method, the four-body interactions arising in models with $2+1$ dimensions and higher, are obtained stroboscopically, through a sequence of two-body interactions with ancillary degrees of freedom. This yields stronger interactions than the ones obtained through pertubative methods, as typically done in previous proposals, and removes an important bottleneck in the road towards experimental realizations. The scheme applies to generic gauge theories with Lie or finite symmetry groups, both Abelian and non-Abelian. As a concrete example, we present the construction of a digital quantum simulator for a $\mathbb{Z}_{3}$ lattice gauge theory with dynamical fermionic matter in $2+1$ dimensions, using ultracold atoms in optical lattices, involving three atomic species, representing the matter, gauge and auxiliary degrees of freedom, that are separated in three different layers. By moving the ancilla atoms with a proper sequence of steps, we show how we can obtain the desired evolution in a clean, controlled way.
\end{abstract}

\maketitle

\tableofcontents

\section{Introduction}

Gauge theories play a very significant role in modern physics. They are responsible, through their special local symmetry - local gauge invariance - of mediating the interactions
between matter particles - either elementary particles in the standard model of particle physics (electromagnetic, strong or weak interactions), or composite particles at lower energy
scales (in  electrodynamics, or some effective, emergent gauge theories). Beside their important role in physics,
gauge theories manifest a
variety of nontrivial physical phenomena and a rich phase diagram. Many of them, especially in the cases of non-Abelian symmetries, are still lacking a complete, analytical understanding and are at the frontier of modern physical research. Open questions include the mechanism of quark confinement, or the mass gap in Yang Mills theories \cite{Wilson,Polyakov1977,Greensite2003}, as well as the phase structure of QCD (quantum chromodynamics) \cite{Kogut2004,Fukushima2011} and especially the search for its exotic phases, as color superconductivity.

Over the years, many approaches have been proposed and applied to the theoretical study of gauge theories. A very prominent and successful one is
lattice gauge theory \cite{Wilson,KogutSusskind,KogutLattice},
which has allowed to prove some basic concepts as well as to calculate numerically, using Monte Carlo methods, parts of the hadronic spectrum \cite{Kogut1983,FLAG2013}.
However, Monte Carlo methods in a Euclidean spacetime cannot approach several problems, as, for example, those which involve fermionic matter with a finite
chemical potential (giving rise to the computationally hard sign problem \cite{Troyer2005}). Another desirable feature is real time evolution in Minkowski spacetime, which is absent when time is imaginary.

Recently, two alternative approaches to revealing the mysteries of lattice gauge theories have been proposed by the quantum information and quantum optics communities. The first suggests using tensor network states \cite{Verstraete2008,Orus2014} to study the ground states, time evolution and phase structure of lattice gauge theories, with both numerical and analytical methods
\cite{Byrnes2002,Banuls2013,Banuls2013a,Rico2014,Tagliacozzo2014,Haegeman2014,Kuhn2014,Buyens2013,Silvi2014,Kuhn2015,Pichler2015,Banuls2015,Zohar2015b,Buyens2015,Zohar2016,Banuls2016,Buyens2016,Zohar2016a}.
The second exploits the diverse variety of optical, atomic and solid-state devices, which are nowadays controllable in experiment, as quantum simulators \cite{Feynman1982,CiracZoller}
of lattice gauge theories \cite{Wiese2013,Zohar2015a}: i.e. specially-tailored quantum systems which mimic the behavior of the quantum theories of interest, serving as playgrounds for the study of otherwise inaccessible physics.
The key issue in quantum simulation of lattice gauge theories is the need to
enforce local gauge symmetry on the simulating systems - a symmetry which these systems do not exhibit explicitly, but is the most important ingredient of the simulated models.

There are two main approaches for quantum simulation. One is analog, where the degrees of freedom and the dynamics of the desired gauge theory are fully or approximately mapped to the simulating system by imposing some external constraints. Another is digital simulation, where the simulating system is evolved stroboscopically by applying a precise sequence of short quantum operations that approximates, to a given precision, the dynamics of the simulated system~\cite{Jane2003a}.

Many quantum simulators have been recently proposed, based, for example, on ultracold atoms in optical lattices, trapped ions or superconducting qubits. These proposals have addressed lattice gauge theories of different levels of complexity, Abelian or non-Abelian, with or without dynamical matter
\cite{Zohar2011,Zohar2012,Banerjee2012,Tagliacozzo2013,Zohar2013,NA,Topological,Rishon2012,TagliacozzoNA,AngMom,ZollerIons,SQC,Dissipation,Kosior2014,Marcos2014,Wiese2014,Pepe2015,Mezzacapo2015,Walter2015,Kasper2015,Laflamme2015,Omjyoti2016,Kasper2016,Dayou2016}. A quantum simulation of the lattice Schwinger model (electrodynamics in $1+1d$) has even been realized experimentally, using trapped ions~\cite{Martinez2016}.

Still, the experimental realization of quantum simulators for lattice gauge theories is very challenging in general. First, many proposals require the use of sophisticated experimental techniques - e.g. Feshbach resonances or single addressability in some ultracold atoms proposals. Second, lattice gauge theories in $2+1$ dimensions and more, involve four-body interactions (the plaquette magnetic interactions) whose implementation is nontrivial. In previous proposals, these are obtained by using perturbation theory and effective Hamiltonian terms, which make them very weak. For example, if one wishes to obtain these interaction terms out of already gauge invariant building blocks, fourth order processes are needed~\cite{AngMom}, whose experimental realization is of course very difficult. This sets a major bottleneck for pushing experiments beyond the simple $1+1d$ case.

Digital quantum simulation may be a way of overcoming this bottleneck. Some previous methods for digital quantum simulators of lattice gauge theories using cold atoms (in particular Rydberg atoms) have  recently been proposed: One \cite{Rydberg} dealt with a simulation of a $U(1)$ pure gauge theory with two possible electric field values; another \cite{Tagliacozzo2013} with a quantum simulation of $U(1)$ and $\mathbb{Z}_N$ gauge magnets, with static charges; and  \cite{TagliacozzoNA} proposed to simulate an $SU(2)$ quantum magnet. Here we provide a general digital construction of lattice gauge theories, including dynamical matter, for any reasonable gauge group.

In a recent work~\cite{ZNsim} we introduced the scheme and its implementation with cold atoms, considering  the particular example of a $\mathbb{Z}_{2}$ lattice gauge theory. In this work we expand, elaborate and generalize the discussion. After reviewing some basic ingredients of lattice gauge theories, we will formulate a general digital construction of a lattice gauge theory. As in previous proposals this is based on a lattice system which includes, in addition to the gauge and matter degrees of freedom, some auxiliary particles that mediate the required interactions and eventually give rise to a dynamics which is equivalent to that of the desired gauge theory. In particular, we are interested in constructing a stroboscopic evolution from small time steps. We will show how to build individual time steps that respect local gauge invariance, so that any error due to the digitization will not break the symmetry of the system. Moreover we will show how all the required  three- and four-body interactions, including the gauge-matter coupling, can be obtained by concatenating simpler two-body interactions between the physical ingredients and the ancillary degrees of freedom. In our general construction, valid for any gauge group $G$ which is either compact or finite, this task is greatly simplified by the use of a mathematical quantum mechanical object called \emph{stator}. We show here how stators, introduced and described in \cite{Reznik2002,ZoharStators} prove to be a very powerful and useful tool in the context of digital lattice gauge theories.

Then, we shall turn into a detailed construction of a particular example: simulating a $\mathbb{Z}_3$ lattice gauge theory in $2+1$ dimensions using ultracold atoms in optical lattices, trapped in a layered structure which allows the ancillary atoms to move and interact with the ``physical" ones \cite{Aguado2008}.
We anticipate that there are strong qualitative differences between the $\mathbb{Z}_3$ implementation and the $\mathbb{Z}_2$ case discussed in~\cite{ZNsim}, as the latter benefits from several simplifications that are no longer available for any $N \geq 2$. Therefore, the $\mathbb{Z}_3$ case takes a considerable step forward with respect to \cite{ZNsim} and constitutes an interesting example containing new ingredients that can be readily extended to higher values of $N$. Readers interested only in the particular $\mathbb{Z}_3$ realization scheme, may skip the general framework and jump to section \ref{appz3}.

Throughout this paper, a summation convention is assumed for double indices. An exception is in the case of irreducible representation indices, whose summation takes place only if
it is explicitly written.

\section{The digitization scheme}
First we shall describe how to construct the dynamics of lattice gauge theories in a digital way, for a generic compact or finite symmetry group, regardless of their experimental feasibility.

\subsection{Mathematical preliminaries}

\subsubsection{Lattice gauge theories: the physical ingredients}

\begin{figure}
  \centering
  % Requires \usepackage{graphicx}
  \includegraphics[width=0.5\columnwidth]{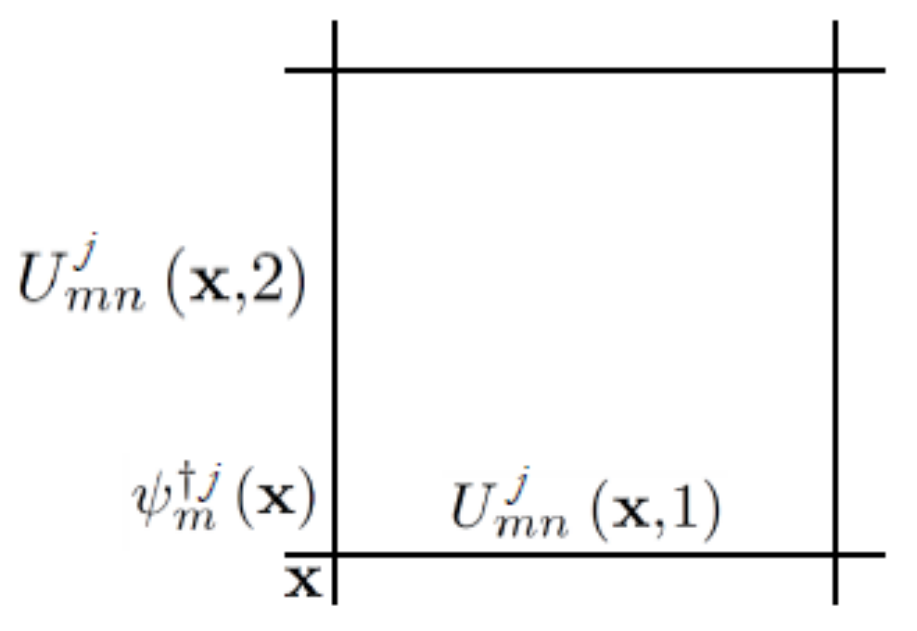}\\
  \caption{A single plaquette. It is labeled after the vertex on the left bottom, $\mathbf{x}$, where the fermionic spinor $\psi^{j\dagger}_m\left(\mathbf{x}\right)$
  resides. The gauge field operators $U^{j}_{mn}\left(\mathbf{x},1\right)$ and $U^{j}_{mn}\left(\mathbf{x},2\right)$ act on the gauge field Hilbert spaces
  of the links $\left(\mathbf{x},1\right)$ and $\left(\mathbf{x},2\right)$ respectively.}
  \label{plaqfig}
\end{figure}

Let $G$ be some group, which may be either a compact Lie or a finite (discrete) group. $G$ has, in general, several irreducible representations $j$. We define a Hilbert space based on
the group $G$, with elements spanned by a basis of the form $\left|j m\right\rangle$, where $j$ is the representation and $m$ identifies state within the
same representation. To explain this, let us introduce a unitary operator $\theta_g$ - the quantum operator responsible for transforming states with respect to the group element $g \in G$.
Such an operator is block-diagonal in the representation, i.e. group transformations leave the representation invariant. Thus, states should be labeled with respect to the representation.
Within a given representation, on the other hand, the states will be mixed by the transformation, and thus we need another quantum number to quantify that. The transformation $\theta_g$ is unitary,
and thus it will be described by the unitary matrices $D^j_{mn}\left(g\right)$, corresponding to the representation $j$ and the group element $g$:
\begin{equation}
\theta_g \left|jm\right\rangle = D^{j}_{nm}\left(g\right)\left|jn\right\rangle.
\end{equation}
These are the generalized Wigner matrices \cite{Rose1995},
\begin{equation}
D^{j}_{mn}\left(g\right) = \left\langle jm\right|\theta_g\left|jn\right\rangle.
\end{equation}

To understand better the states $\left|jm\right\rangle$, consider, for example, $G=SU\left(2\right)$. Then $j$ labels the total angular momentum of a state and $m$ its $z$ component; Thus, in general,
$m$ may be thought of as a set of quantum numbers or indices, corresponding to the eigenvalues of a maximal set of mutually commuting operators that also commute with the representation operator of the
group $G$. For $SU(2)$, the ``representation operator" is $\mathbf{J}^2$, and the maximal set of mutually commuting operators contains a single operator, an angular momentum component,
usually chosen to be $J_z$. For $SU(3)$, on the other hand, $m$ will be a set of two quantum numbers, the hypercharge and isospin, eigenvalues of two operators which commute with one
another as well as with the Casimir operator which is block diagonal in the representation.

Lattice gauge theories involve matter degrees of freedom (mostly, as in our case, fermions), residing on the vertices $\mathbf{x}\in \mathbb{Z}^d$ of a $d$ dimensional spatial lattice,
and gauge fields residing on its links, labeled by a pair $\left(\mathbf{x},k\right)$ of a vertex $\mathbf{x}$ from which they emanate and a (positive) direction $k=1...d$ in which
they emanate. We denote by $\hat{\mathbf{k}}$ the unit (lattice) vector pointing in the $k$th direction.
We will restrict our discussion to $d=2$ - i.e. lattice gauge theories in $2+1$ dimensions - but it can be straightforwardly generalized to other dimensions (see Fig. \ref{plaqfig}).

The states and Hamiltonians of lattice gauge theories are locally invariant under some gauge group $G$. By locally we mean that the state of the system is invariant under a special type
of group transformations (local gauge transformations) which depend on different transformation parameters - or group elements - for different positions $\mathbf{x}$. We will first
describe the physical degrees of freedom and their local Hilbert spaces, which will allow us then to explain these special symmetry properties.

First, let us consider the type of Hilbert spaces which will be used for our lattice fermionic matter field.
In a fermionic Fock space, we may define fermionic creation operators $\psi^{j\dagger}_m$ that create, from the Fock vacuum, states which transform under $G$ as $\left|jm\right\rangle$. In operator terms, this reads
\begin{equation}
\theta_g^{\psi} \psi^{j\dagger}_n \theta_g^{\psi\dagger} = \psi^{j\dagger}_m D^{j}_{mn}\left(g\right).
\end{equation}
Then, for a fixed $j$, the set of operators $\psi^{j\dagger}_m$ forms a spinor of the $j$ representation. Such spinors, as we shall see, describe the matter degrees of freedom in a
lattice gauge theory.

If $G$ is a compact Lie group, $\theta_g$ may be written as
\begin{equation}
\theta_g = e^{i \phi_a Q_a}
\end{equation}
where $\phi_a$ are the group parameters corresponding to $G$, and $Q_a$ is the charge at the vertex. $Q_{a}$ satisfies the group's algebra
\begin{equation}
\left[Q_a,Q_b\right]=if_{abc}Q_c
\end{equation}
where $f_{abc}$ are the structure constants of $G$. After defining $T^j_a$, the $j$ matrix representation of the $a$ generator,
\begin{equation}
\left[T_a,T_b\right]=if_{abc}T_c
\end{equation}
we can explicitly define
\begin{equation}
Q_a = \psi^{j\dag}_m \left(T^{j}_a\right)_{mn} \psi^j_n.
\end{equation}

On the vertices of our lattice, we place fermionic spinors $\psi^{j\dagger}_m$, belonging to some irreducible representation $j$ of the gauge group $G$. We will omit the index $j$, assuming that we only
include spinors of one fixed representation, but in general one may include more than one fermionic representation. In most of the ``conventional" lattice gauge theories associated
with Lie groups, one picks the fundamental representation for the matter spinors (e.g. $j=1/2$ for $SU(2)$).

We will choose, for convenience, to work with staggered fermions \cite{Susskind1977,Zohar2015} - i.e., fermions occupying even vertices correspond to particles, and absences in odd vertices to anti-particles. If there exists a continuum limit,
such two spinors will be united to a single Dirac spinor. The charge of such fermions is updated to
\begin{equation}
\theta_g = e^{i \phi_a Q_a} \left(\det\left(D^j\left(g^{-1}\right)\right)\right)^N
\end{equation}
where $N=0$ for even vertices and $N=1$ for odd ones.

The fermionic vacuum, called the "Dirac sea" $\left|D\right\rangle$, is a state in which all the odd vertices are fully occupied and the even ones are empty - this corresponds to having neither particles nor
anti-particles in the system. We will define the transformation of the empty Fock vacuum $\left|\Omega\right\rangle$ as
\begin{equation}
\theta_g\left(\mathbf{x}\right)\left|\Omega\right\rangle = \left\{
                                                                              \begin{array}{ll}
                                                                                \left|\Omega\right\rangle, & \hbox{$\mathbf{x}$ even;} \\
                                                                                \det\left(D\left(g^{-1}\right)\right)\left|\Omega\right\rangle, & \hbox{$\mathbf{x}$ odd.}
                                                                              \end{array}
                                                                            \right.
\end{equation}
Then, since
\begin{equation}
\theta_g \underset{m}{\prod} \psi^{\dagger}_m \theta_g^{\dagger} = \det\left(D\left(g\right)\right) \underset{m}{\prod} \psi^{\dagger}_m
\end{equation}
\cite{Zohar2015}
we get that the Dirac sea $\left|D\right\rangle$ is invariant under the fermionic transformations.

The gauge fields are represented in another way \cite{Zohar2015,Zohar2016}.
On each link of the lattice, there exists a Hilbert space of a second type, spanned either by group element states or representation states \cite{Zohar2015}, which we shall explain now.

The \emph{group element basis} consists of states attached to group elements, $\left|g\right\rangle$, on which one may
act with the following unitary transformations
\begin{equation}
\begin{aligned}
&\Theta_g\left|h\right\rangle = \left|hg^{-1}\right\rangle \\
&\tilde{\Theta}_g\left|h\right\rangle = \left|g^{-1}h\right\rangle.
\end{aligned}
\end{equation}

We define a matrix of operators (in an "internal" \emph{group, gauge or matrix space}),
$U^{j}_{mn}$, whose elements are operators in Hilbert space, by
\begin{equation}
U^{j}_{mn} = \int dg D_{mn}^{j}\left(g\right) \left|g\right\rangle\left\langle g\right| .
\label{Ujmndef}
\end{equation}
This is a ``group element" operator.
Under $\Theta_g,\tilde{\Theta}_g$, it respects the transformation rules
\begin{equation}
\begin{aligned}
&\Theta_g U^{j}_{mn} \Theta^{\dagger}_g = U^{j}_{mn'} D^{j}_{n'n} \left(g\right)\\
&\tilde\Theta_g U^{j}_{mn} \tilde\Theta_g^{\dagger} = D^{j}_{mm'} \left(g\right) U^{j}_{m'n}.
\end{aligned}
\end{equation}

If $G$ is a compact Lie group, we may expand
\begin{equation}
U^{j} = e^{i \hat\phi_a T^j_a}
\label{UexpT}
\end{equation}
where $\hat\phi_a$ are \emph{operator} group parameters corresponding to $g$.
The transformations may also be expanded in terms of generators: define the right and left generators, $R_a,L_a$, such that
\begin{equation}
\begin{aligned}
&\left[R_a,R_b\right]=if_{abc}R_c \\
&\left[L_a,L_b\right]=-if_{abc}L_c \\
&\left[L_a,R_b\right]=0 \\
&\left[R_a,U^j_{mn}\right]=U^{j}_{mn'}\left(T^j_a\right)_{n'n} \\
&\left[L_a,U^j_{mn}\right]=\left(T^j_a\right)_{mm'}U^{j}_{m'n} \\
&\mathbf{J}^2 \equiv R_a R_a = L_a L_a
\end{aligned}
\label{algebra}
\end{equation}
and then
\begin{equation}
\begin{aligned}
&\Theta_g = e^{i \phi_a R_a} \\
&\tilde\Theta_g = e^{i \phi_a L_a}.
\end{aligned}
\end{equation}
Note that $U^j$ does not commute with the generators, and thus the representation number $j$ is a dynamical quantity.

The group parameters $\hat{\phi}_a$ on the links are the (generally) non-Abelian, color components of the vector potential. Thus one may call the group element basis "magnetic basis"
as well. The generators, on the other hand, are some non-Abelian extension of ``conjugate" degrees of freedom to the vector potential, and thus stand for the (generally non-Abelian, right and left)
electric fields. This automatically explains the need for the \emph{representation basis}, $\left|jmn\right\rangle$. Every state in this basis is composed out of
three quantum numbers (or sets of which): the representation $j$, and identifiers within the representation $m,n$. This is similar to the $m$ in $\left|jm\right\rangle$, however, here we
have two sets in general, one corresponding to left transformations and the other to right ones (which are generally different as $G$ may be non-Abelian). These will be the eigenvalues
of the operators in the maximally mutual commuting sets, one for the left ($m$) and one for the right ($n$), as may be deduced from the algebra (\ref{algebra}). For $SU(2)$, for example,
we can choose $m$ as the eigenvalue of $L_z$, and $n$ as that of $R_z$.

The motivation for introducing the representation basis is very clear for compact Lie groups, but one can do it formally for finite groups as well. In both cases, the change
of basis is defined by the Wigner matrices,
\begin{equation}
\left\langle g | jmn \right\rangle = \sqrt{\frac{\dim\left(j\right)}{\left|G\right|}}D^{j}_{mn}\left(g\right)
\end{equation}
where $\left|G\right|$ is the order of $G$. A singlet state is given by $\left|000\right\rangle$; then,   it is possible to show \cite{Zohar2015} that
\begin{equation}
\left|jmn\right\rangle = \sqrt{\dim\left(j\right)}U^{j}_{mn}\left|000\right\rangle.
\end{equation}

Although the matrix elements of $U^j_{mn}$ are Hilbert space operators, they fulfill a very special property: they commute with one another, as a direct consequence of (\ref{Ujmndef}): the matrix elements $U^j_{mn}$ are all diagonalized in the same basis and thus they commute. Thus, when calculating, one may treat the elements
of $U^j_{mn}$ as numbers for many purposes. In particular, we may define ``group space operations" or ``matrix operations", which involve acting only on the matrix indices. For example,
when we talk about the trace of a $U^j$ operator, or of an operator composed of such operators, it is a trace in the matrix space (or group space), which is still an operator in Hilbert
space: the sum of Hilbert space operators on the diagonal of $U^j$. One has thus to be careful with the notion of hermitian conjugation, and pay attention to whether it is performed in
the matrix space, or in the Hilbert space. The following relation is very helpful for that purpose:
\begin{equation}
\begin{aligned}
&\left(U^j_{mn}\right)^{\dagger} = \int dg \left|g\right\rangle\left\langle g \right| \overline{D}^{j}_{mn}\left(g\right) = \\
&\int dg \left|g\right\rangle\left\langle g \right| D^{j\dagger}_{nm}\left(g\right) = \left(U^{j\dagger}\right)_{nm}
\end{aligned}
\label{daggerdagger}
\end{equation}
i.e., the conjugate transpose in Hilbert space of a matrix element is equal to the matrix element in the transposed position of the hermitian conjugation in matrix space.
We may also define matrix functions which are ``blind" to the Hilbert space structure: for example, define
\begin{equation}
Z^j_{mn} = -i \left(\text{log}_{\text{mat}}\left(U^j\right)\right)_{mn}
\end{equation}
where $\text{log}_{\text{mat}}$ means that the logarithm is taken only in matrix space (which is well defined thanks to the commutativity, in Hilbert space, of the matrix elements of $U$).

Thus, as a consequence of (\ref{daggerdagger}),
\begin{equation}
\left(Z^j_{mn}\right)^{\dagger} = Z^j_{nm}
\label{Zdagger}
\end{equation}
where the hermitian conjugation is taken \emph{in the Hilbert space}. If $G$ is a compact group, we simply obtain that
\begin{equation}
Z^j_{mn} = \hat{\phi}^a (T^j_a)_{mn}.
\label{ZComp}
\end{equation}

From now on, we shall omit representation indices from the $U^{j}_{mn}$ in case they belong to the fundamental representation of the group.

\subsubsection{Lattice local gauge invariance}

With the definitions made above at hand, we can finally define a local gauge transformation: this is a transformation which acts on all the Hilbert space intersecting at the vertex $\mathbf{x}$ - i.e.,
the fermions at the vertex and all the links starting and ending there - and depends on a group element which may differ as a function of the position, $g\left(\mathbf{x}\right)$.
The gauge transformation is defined as
\begin{equation}
\hat \Theta_g\!\left(\mathbf{x}\right) \!=\! \tilde{\Theta}_g\!\left(\mathbf{x},1\right) \! \tilde{\Theta}_g\!\left(\mathbf{x},2\right) \!
\Theta^{\dagger}_g\!\left(\mathbf{x} \!-\! \hat{1},1\right) \! \Theta^{\dagger}_g\!\left(\mathbf{x} \!-\! \hat{2},2\right) \! \theta^{\dagger}_g\!\left(\mathbf{x}\right).
\label{gtran}
\end{equation}

A gauge invariant state $\left|\psi\right\rangle$ is invariant under such a transformation \footnote{Up to a phase, representing static charges which we shall not treat in this work.},
\begin{equation}
\hat \Theta_{g\left(\mathbf{x}\right)}\left(\mathbf{x}\right) \left|\psi\right\rangle =  \left|\psi\right\rangle ,\quad \forall \mathbf{x} .
\end{equation}

If $G$ is a compact Lie group, we may define generators $G_a\left(\mathbf{x}\right)$, satisfying the group algebra, such that
\begin{equation}
\begin{aligned}
\left[G_a,G_b\right] \!=  &if_{abc}G_c ,\\
G_a\!\left(\mathbf{x}\right) \!= & L_a\!\left(\mathbf{x},1\right) \!+\! L_a\!\left(\mathbf{x},2\right)\\ & \!-\! R_a\!\left(\mathbf{x} \!-\! \hat{1},1\right) \!-\! R_a\!\left(\mathbf{x} \!-\! \hat{2},2\right) \!-\! Q_a\!\left(\mathbf{x}\right).
\end{aligned}
\end{equation}
Then, the equation
\begin{equation}
G_a \left|\psi\right\rangle = 0, \quad \forall \mathbf{x},a
\end{equation}
is satisfied for a gauge invariant $\left|\psi\right\rangle$ and forms a lattice \emph{Gauss Law}, which gives a better intuition for identifying the right and left generators
as electric fields.

The ``global singlet state"
\begin{equation}
\left|0\right\rangle \equiv \left|D\right\rangle \underset{\text{links}}{\bigotimes}\left|000\right\rangle
\label{globsing}
\end{equation}
is invariant under gauge transformations (\ref{gtran}), and any other gauge invariant product may be obtained by acting on it with a product of gauge invariant operators. There are
several types of such operators:
\begin{enumerate}
  \item \noindent Traces of products of $U$ and $U^{\dagger}$ operators along a closed path. The simplest is the plaquette operator.
  \begin{equation}
  {\rm Tr} \!
  \left(U\!\left(\mathbf{x},1\right)\!
        U\!\left(\mathbf{x}\!+\!\hat{1},2\right)\!
        U^{\dagger}\!\left(\mathbf{x}\!+\!\hat{2},1\right)\!
        U^{\dagger}\!\left(\mathbf{x},2\right)\right)
  \end{equation}
  \item Products of $U$ and $U^{\dagger}$ operators along a line, with fermionic operators at the egdes. The simplest is the link interaction,
  \begin{equation}
  \psi^{\dagger}_m\left(\mathbf{x}\right)U_{mn}\left(\mathbf{x},k\right)\psi_n\left(\mathbf{x}+\hat{\mathbf{k}}\right).
  \end{equation}
  \item Local fermionic group scalars:\\ ${ \psi^{\dagger}_m\!\left(\mathbf{x}\right)\!\psi_m\!\left(\mathbf{x}\right) \!\equiv\! \psi^{\dagger}\!\left(\mathbf{x}\right)\!\psi\!\left(\mathbf{x}\right) }$
  \item Gauge field operators which are diagonal in the representation basis. The simplest case is the representation operator,
  \begin{equation}
  \Pi\left(\mathbf{x},k\right) = \underset{j}{\sum}f\left(j\right)\left|jmn\right\rangle\left\langle jmn\right|.
  \end{equation}
\end{enumerate}

A reasonable Hamiltonian for lattice gauge involves such terms. In general, such a Hamiltonian will involve four terms:
\begin{enumerate}
  \item The \emph{Electric Hamiltonian},
  \begin{equation}
  H_E = \lambda_E \underset{\mathbf{x},k}{\sum}\Pi\left(\mathbf{x},k\right).
  \end{equation}
  It is called the electric Hamiltonian, because in the $SU(N)$ case the quadratic Casimir operators $\mathbf{J}^2$ can take the role of $\Pi$ and then this part is just
  a sum over the square of the electric field everywhere, i.e. the electric energy.
  \item The \emph{Magnetic Hamiltonian},
  \begin{align}
  &H_B =  \underset{\mathbf{x}}{\sum}H_B\left(\mathbf{x}\right) =\\
  &\lambda_B \! \underset{\mathbf{x}}{\sum}
  \text{Tr}
  \left(U\!\left(\mathbf{x},1\right)\!
        U\!\left(\mathbf{x} \!+\! \hat{1},2\right)\!
        U^{\dagger}\!\left(\mathbf{x} \!+\! \hat{2},1\right)\!
        U^{\dagger}\!\left(\mathbf{x},2\right)\right) \!+\! H.c. \nonumber
  \end{align}
  Here, in the case of a Lie group, we could obtain in the continuum limit the magnetic energy term - e.g. sum of the magnetic field squared for QED ($G=U(1)$).
  \end{enumerate}
  These first two parts
  describe only the gauge field, and in the case of compact Lie groups they both add up to the \emph{Kogut-Susskind Hamiltonian} $H_{KS} = H_E+H_B$, the Hamiltonian of a lattice
  Yang-Mills theory \cite{KogutSusskind,KogutLattice}.
  \begin{enumerate}
  \item[3.] The fermionic mass Hamiltonian,
  \begin{equation}
  H_M = M\underset{\mathbf{x}}{\sum}(-1)^{x_1+x_2}\psi^{\dagger}\left(\mathbf{x}\right)\psi\big(\mathbf{x}\big)
  \end{equation}
  in which the alternating signs stand for the staggering of fermions: particles on even sites, anti-particles on odd ones \cite{Susskind1977,Zohar2015}.
  \item[4.] The gauge-matter interaction,
  \begin{equation}
  \begin{aligned}
  H_{GM} =& \underset{\mathbf{x},k}{\sum}H_{GM}\left(\mathbf{x},k\right)=\\
  &\lambda_{GM}
  \underset{\mathbf{x},k}{\sum}
  \psi^{\dagger}_m\! \left(\mathbf{x}\right) \! U_{mn}\!\left(\mathbf{x},k\right) \! \psi_n\!\big(\mathbf{x}+\hat{\mathbf{k}}\big) \!+\! H.c.
  \end{aligned}
  \end{equation}
\end{enumerate}

In this work, we will discuss a digital implementation of the total Hamiltonian,
\begin{equation}
\label{Eq:Hamilt}
H = H_{E}+H_{B}+H_{M}+H_{GM}.
\end{equation}

\subsection{The digitization}
\label{SecDigit}

Under the action of the Hamiltonian $H$, the system evolution is described by the unitary operator ${ \mathfrak{U}(t) = e^{-i H t} }$.
The total Hamiltonian $H$ is, however, hard to implement as a whole in a cold atomic system, due to the interacting parts $H_B$ and $H_{GM}$.
 Instead, we can easily implement
 separate parts of the Hamiltonian if we consider them individually.
  In our particular case, as we shall see, the terms whose evolution can be implemented independently are:
  ${ H_E }$, ${ H_M }$, $H_{Be}$, ${ H_{Bo} }$, ${ H_{GM,eh} }$, ${ H_{GM,ev} }$, ${ H_{GM,oh} }$ and ${ H_{GM,ov} }$,
  where
    \begin{equation}
	\label{Eq:Z3SimulatedHamiltonianEvenPlaquette}
	H_{Be} =  \sum_{\mathbf{x}\text{ even}}  H_B\left(\mathbf{x}\right)
\end{equation}
involves a sum only over even plaquettes and similarly $H_{Bo}$ involves a sum only over odd ones;
\begin{equation}
	\label{Eq:Z3SimulatedHamiltonianEvenPlaquette}
	H_{GM,eh} =  \sum_{\mathbf{x}\text{ even}} H_{GM}\left(\mathbf{x},1\right)
\end{equation}
describes the gauge-matter interaction on horizontal links originating from even sites and ${ H_{GM,ev} }$, ${ H_{GM,oh} }$, ${ H_{GM,ov} }$ are defined in a similar way.

We can then use each single term to evolve the system for a very short time $\tau$,
 i.e. we can separately realize the unitary operators ${ W_E= e^{-i H_E \tau} }$,
  $ W_M = e^{-i H_M \tau} $, ${ W_{Be} = e^{-i H_{Be} \tau} }$,
  $W_{eh}=e^{-i H_{GM,eh}\tau}$ etc. Then, by Trotter formula we have ${ e^{-i \Sigma_j H_j t} \!=\! \lim_{M\rightarrow \infty}\! \big( \mathbf{\Pi}_j \; e^{-i H_j t/M} \big)^M  \!\equiv\! \lim_{M\rightarrow \infty}\! \big( W(t/M) \big)^M }$ \cite{Trotter1959a,Suzuki1985a,Bhatia1996}
and we see, putting $\tau=t/M$, that we can approximate the total evolution with a specific sequence of short evolutions according to each of the pieces $H_j$ \cite{LLoyd1996a,Jane2003a}. Further details on the digitization of the evolution will be given in a later section.

\subsubsection{Stators}
A useful ingredient in our digital formulation of lattice gauge theories is a mathematical object called \emph{stator} \cite{Reznik2002} - ``state-operator". It is somewhat a ``mixture" of an operator and a state, living in a product Hilbert space. We shall briefly describe some of its most relevant properties; For further discussion and generalizations, the reader may refer to \cite{ZoharStators}.

Consider two Hilbert spaces, $\mathcal{H}_A$ and $\mathcal{H}_B$. Denote the set of operators acting on a Hilbert space $\mathcal{H}$ by $\mathcal{O}\left(\mathcal{H}\right)$. Then, a stator $S \in \mathcal{O}\left(\mathcal{H}_A\right) \times \mathcal{H}_B$ will be the result of acting with a unitary $\mathcal{U}_{AB} \in \mathcal{O}\left(\mathcal{H}_A \times \mathcal{H}_B\right)$
on some initial state $\left|in\right\rangle \in \mathcal{H}_B$:
\begin{equation}
S = \mathcal{U}_{AB} \left|in_B\right\rangle \in  \mathcal{O}\left(\mathcal{H}_A\right)\times \mathcal{H}_B.
\end{equation}
It may be written as an expansion of the form
\begin{equation}
S = \underset{i}{\sum}M_i \otimes \left|i_B\right\rangle
\end{equation}
where $M_i \in \mathcal{O}\left(\mathcal{H}_A\right)$ are Kraus operators satisfying $\underset{i}{\sum}M_i^{\dagger}M_i = \mathbf{}{1}_A$, and $\left|i_B\right\rangle \in \mathcal{H}_B$. Mathematically speaking, $S$ is an isometry that maps a state $\ket{\psi_{A}}$ of the physical Hilbert space $\mathcal{H}_{A}$ to the tensor product $\mathcal{U}_{AB} \left( \ket{\psi_{A}} \otimes \left|in_B\right\rangle \right)$, living in $\mathcal{H}_A \times \mathcal{H}_B$.

We say that the operators $\Theta_A \in \mathcal{O}\left(\mathcal{H}_A\right)$ and ${ \Theta_B \in \mathcal{O}\left(\mathcal{H}_B\right) }$ are a pair of $S$-eigenoperators, if the following holds:
\begin{equation}
\label{Eq:StatorRel}
\Theta_B S = S \Theta_A
\end{equation}

Relation \eqref{Eq:StatorRel} is useful for digital schemes in the following way. Suppose that $\mathcal{H}_A$ describes our ``physical system" and
$\mathcal{H}_B$ describes some auxiliary degree of freedom. We have an initial product state $\left|\psi_A\right\rangle \left|in_B\right\rangle$, and
we wish to evolve the physical state $\left|\psi_A\right\rangle$ for some time $t$ with a Hamiltonian $H \in \mathcal{O}\left(\mathcal{H}_A\right)$.
 If there are a stator $S$ and a Hermitian operator $H' \in \mathcal{O}\left(\mathcal{H}_B\right)$ such that
\begin{equation}
H' S = S H
\end{equation}
we get, as well, that
\begin{equation}
e^{-i H' t} S = S e^{-i H t}.
\end{equation}
Therefore we can obtain, effectively, a time evolution of the physical state by creating a stator of the physical and auxiliary ingredients, and then acting on the control:
\begin{equation}
e^{-i H' t} \mathcal{U}_{AB} \! \left|\psi_A\right\rangle \! \left|in_B\right\rangle \!=\!
e^{-i H' t} S \! \left|\psi_A\right\rangle \!=\!
S e^{-i H t} \! \left|\psi_A\right\rangle.
\end{equation}
I.e., we first let the physical and the control systems interact such that $U_{AB}$ is generated, and then turn on a Hamiltonian $H_B$ for the control for a time period $t$.
Then, the stator is ready for the next step, or can be undone by letting the two systems interact again to realize $U^{\dagger}_{AB}$. In the latter case we get
\begin{equation}
\mathcal{U}^{\dagger}_{AB} e^{-i H' t} \mathcal{U}_{AB} \left|\psi_A\right\rangle \left|in_B\right\rangle =
\left|in_B\right\rangle \otimes e^{-i H t} \left|\psi_A\right\rangle
\end{equation}
i.e. we end with a product state, where the physical state has evolved according to the desired Hamiltonian $H$ and the auxiliary system is back in its initial state.

In a similar way one can, for example, create a stator which connects several physical degrees of freedom together, and then obtain effectively a many body interaction among them. This is what we will do in order to obtain the plaquette interactions. For that, we shall introduce a special kind of stator, called \emph{Group element stator}:
\begin{equation}
S = \int dg \left|g_A\right\rangle  \left\langle g_A\right| \otimes  \left|g_B\right\rangle.
\end{equation}
Following \eq{Ujmndef}, we get the eigenoperator relations
\begin{equation}
\left(U^j_{mn}\right)_B S = S \left(U^j_{mn}\right)_A
\end{equation}
and
\begin{equation}
\left(U^{j\dagger}_{mn}\right)_B S = S \left(U^{j\dagger}_{mn}\right)_A.
\end{equation}
Consider now a single plaquette in a lattice gauge theory, whose links are labeled by $1-4$, counterclockwise from $\left(\mathbf{x},1\right)$ to $\left(\mathbf{x},2\right)$, and have local Hilbert spaces $\left\{\mathcal{H}_i\right\}_{i=1}^4$
describing the gauge degrees of freedom. We introduce an auxiliary degree of freedom, serving as the control, in the middle of the plaquette and we assign to it a similar Hilbert space $\widetilde{\mathcal{H}}$.

We define the unitary operator (interaction) $\mathcal{U}_i$, which creates a group element stator for the link $i$ and the control.
For example, if the initial state of the control is $\left|\widetilde{in}\right\rangle=\left|\widetilde{e}\right\rangle$ (the group element state corresponding to the identity element), we have
\begin{equation}
\mathcal{U}_i = \int dg \left|g_i\right\rangle  \left\langle g_i\right| \otimes \tilde{\Theta}^{\dagger}_g
\end{equation}
but this is only one possible choice.

We define then a ``plaquette stator" which is the result of acting on the initial state of the control with the following sequence
\begin{equation}
S_{\square} = \mathcal{U}_{\square} \left|\widetilde{in}\right\rangle \equiv   \mathcal{U}_1  \mathcal{U}_2 \mathcal{U}^{\dagger}_3 \mathcal{U}^{\dagger}_4\left|\widetilde{in}\right\rangle,
\end{equation}
i.e. a sequence of four two-body interactions creates a plaquette stator for four physical degrees of freedom and the control. $S_{\square}$ satisfies the eigenoperator relation
\begin{equation}
\text{Tr}\left(\widetilde{U^j}+\widetilde{U}^{j\dagger}\right) S_{\square} = S_{\square} \text{Tr}\left(U^j_1 U^j_2 U^{j\dagger}_3 U^{j\dagger}_4+ H.c.\right)
\label{eigrelplaq}
\end{equation}
which, as we shall show next, directly gives rise to the magnetic plaquette interaction $H_B$.

Note that this is similar to procedures carried out in previous works, where controlled rotations were utilized for the purpose of obtaining a four body interactions for particular lattice gauge theories (compare for example the plaquette operator $\mathcal{U}_1  \mathcal{U}_2 \mathcal{U}^{\dagger}_3 \mathcal{U}^{\dagger}_4$ with eq. (2) in \cite{Rydberg}, or eq. (58) in \cite{Tagliacozzo2013}). Here, however, we utilize the stator formalism which allows to present things in a more "natural" way,  and to generalize the results to more complicated gauge groups including dynamical fermionic matter as well, as we shall show below.

\subsubsection{The plaquette interactions}

Using the stators defined above, we can obtain the magnetic Hamiltonian $H_B$ in a digital way. What is required, basically, is to create the stator $S_{\square}$ for each plaquette and then act locally on the control with some Hamiltonian $\tilde{H}_{B}$ that yields $H_B$ through the eigenoperator relation \eqref{eigrelplaq}. Therefore, for each plaquette $\mathbf{x}$,
we need to act with the control Hamiltonian
\begin{equation}
\widetilde{H}_B\left(\mathbf{x}\right) = \lambda_B \text{Tr}\left(\widetilde{U^j}\left(\mathbf{x}\right)+\widetilde{U}^{j\dagger}\left(\mathbf{x}\right)\right).
\end{equation}

The sequence
\begin{equation}
\mathcal{U}_p\left(\mathbf{x}\right) = \mathcal{U}^{\dagger}_{\square} \left(\mathbf{x}\right) e^{-i \widetilde{H}_B\left(\mathbf{x}\right)\tau} \mathcal{U}_{\square} \left(\mathbf{x}\right)
\end{equation}
is the unitary operation required for the creation of the plaquette interaction for a single plaquette $\mathbf{x}$, i.e.
\begin{equation}
\mathcal{U}_p\left(\mathbf{x}\right) \left|\psi_{1234}(\mathbf{x})\right\rangle |\widetilde{in}\rangle = |\widetilde{in}\rangle e^{-i H_B\left(\mathbf{x}\right)\tau} \left|\psi_{1234}(\mathbf{x})\right\rangle
\end{equation}
where $\left|\psi_{1234}(\mathbf{x})\right\rangle$ is the initial state of the four links around the plaquette $\mathbf{x}$.

Since $\left[H_B\left(\mathbf{x}\right),H_B\left(\mathbf{y}\right)\right]=0$ for all $\mathbf{x}$ and $\mathbf{y}$, the evolution generated by the global magnetic Hamiltonian $H_B$ is exactly equivalent to the evolution obtained by combining single-plaquette operations $\mathcal{U}_p\left(\mathbf{x}\right)$ across the whole lattice. In mathematical language, we have ${ e^{-i H_{B} \tau} = \prod_{\mathbf{x}} \mathcal{U}_p\left(\mathbf{x}\right) }$. Moreover, different plaquettes can be evolved in parallel or sequentially, with both options leading to the same exact physics. To speed up the simulation, the parallel option is clearly more practical. However, we cannot create the plaquette interactions for all the plaquettes at once, since every link is shared by two plaquettes and thus has to belong to two stators simultaneously, but this is impossible. Then, the fastest way is to realize the $H_B$ evolution in two steps, one for each parity of the plaquettes: for example, we can evolve the even plaquettes first (in parallel), and then the odd ones. In fact, we can even use the same control atoms for both steps, if we are able to move them from even to odd plaquettes.

In summary, the magnetic Hamiltonian $H_{B}$ can be realized as follows. We start with one control atoms placed in the center of each even plaquette and apply the following sequence of operations:
 \begin{enumerate}
   \item Create the stators for even links:
        \begin{enumerate}
        \item  Move all the controls simultaneously to the links below them (even horizontal links), to interact with the gauge field and create the unitary operation
        \begin{equation}
        \mathcal{U}_{1e}\equiv\underset{\mathbf{x}\text{ even}}{\prod}\mathcal{U}_1\left(\mathbf{x}\right).
        \label{U1edef}
        \end{equation}
        Bring the controls back to their "rest position" in the center.
        \item Repeat a similar process with the links on the right, to create
        \begin{equation}
        \mathcal{U}_{2e}\equiv\underset{\mathbf{x}\text{ even}}{\prod}\mathcal{U}_2\left(\mathbf{x}\right).
        \label{U2edef}
        \end{equation}
        \item Proceed with the links above the center, for
        \begin{equation}
        \mathcal{U}^{\dagger}_{3e}\equiv\underset{\mathbf{x}\text{ even}}{\prod}\mathcal{U}^{\dagger}_3\left(\mathbf{x}\right).
        \label{U3edef}
        \end{equation}
        \item Conclude with the last remaining links, and obtain
        \begin{equation}
        \mathcal{U}^{\dagger}_{4e}\equiv\underset{\mathbf{x}\text{ even}}{\prod}\mathcal{U}^{\dagger}_4\left(\mathbf{x}\right).
        \label{U4edef}
        \end{equation}
        \end{enumerate}
   Steps (a)-(d) result in the creation of
   \begin{equation}
   \mathcal{U}_{pe} = \mathcal{U}^{\dagger}_{4e} \mathcal{U}^{\dagger}_{3e} \mathcal{U}_{2e}  \mathcal{U}_{1e} = \underset{\mathbf{x}\text{ even}}{\prod}\mathcal{U}_p\left(\mathbf{x}\right)
   \label{Upedef}
   \end{equation}
   whose action on the initial state of the controls creates a product of the even plaquette stators. The order of the steps (a)-(d) is important, for the order of elements
   in the product obtained in the auxiliary state; if the group is Abelian, on the other hand, this order plays no role.
   \item Turn on the Hamiltonian $\widetilde{H}_{B,e} = \underset{\mathbf{x}\text{ even}}{\sum}\widetilde{H}_B\left(\mathbf{x}\right)$ and let the system evolve for time $\tau$,
    resulting in
   \begin{equation}
   \widetilde{V}_{Be} = e^{-i \widetilde{H}_{B,e} \tau}.
   \label{VBedef}
   \end{equation}
   \item Undo the stators, by a process similar to step 1, but with the inverse interactions - i.e., create $\mathcal{U}^{\dagger}_{pe}$.
 \end{enumerate}
These steps are applied to a state $\left|\psi\left(t_i\right)\right\rangle |\widetilde{in}\rangle$, where
 $\left|\psi\left(t_i\right)\right\rangle$ is the physical state at time $t_i$ and $|\widetilde{in}\rangle$ is the initial state of all the controls. Thanks to the stator relation \eqref{eigrelplaq}, the final result is
 \begin{equation}
 \begin{aligned}
 \left|\psi\!\left(t_i \!+\! \tau\right)\right\rangle\! |\widetilde{in}\rangle \!\equiv\! \mathcal{U}^{\dagger}_{pe} \widetilde{V}_{Be} \mathcal{U}^{\dagger}_{pe}
 \left|\psi\!\left(t_i\right)\right\rangle\! |\widetilde{in}\rangle \!=\! W_{Be} \! \left|\psi\!\left(t_i\right)\right\rangle\! |\widetilde{in}\rangle.
 \end{aligned}
 \end{equation}
 where
 \begin{equation}
 W_{Be} = e^{-i \underset{\mathbf{x}\text{ even}}{\sum}H_B\left(\mathbf{x}\right)\tau}.
 \label{WBedef}
 \end{equation}

 Similarly, one can act on the odd plaquettes, to obtain the time evolution $ W_{Bo}$. For that, the control atoms should first be moved to the center of the odd plaquettes.

\subsubsection{The gauge-matter interactions}

After having achieved the four-body plaquette interactions as a sequence of two-body interactions, we will proceed to obtain the three-body interactions of the matter with the gauge field in a similar way.
The type of interactions we are interested in are those of $H_{GM}$, involving f the gauge field on a link and the fermions on its side. We start by analyzing a single
link, emanating from the vertex $\mathbf{x}$ in the $k$th direction. The corresponding Hamiltonian is
\begin{equation}
H_{GM}\!\left(\mathbf{x},k\right) \! =\! \lambda_{GM} \psi^{\dagger}_m\!\left(\mathbf{x}\right) \! U_{mn}\!\left(\mathbf{x},k\right) \! \psi_n\!\big(\mathbf{x} \!+\! \hat{\mathbf{k}}\big) \!+\! H.c.
\end{equation}

 Thanks to \eqref{Zdagger}, we know that
 $Z_{mn}\left(\mathbf{x},k\right)\psi^{\dagger}_m\left(\mathbf{x}\right)\psi_n\left(\mathbf{x}\right)$ is Hermitian, and
 \begin{equation}
 \mathcal{U}_W\left(\mathbf{x},k\right) = e^{i Z_{mn}\left(\mathbf{x},k\right)\psi^{\dagger}_m\left(\mathbf{x}\right)\psi_n\left(\mathbf{x}\right)}
 \end{equation}
 is unitary. If $G$ is a compact Lie group, from eq. (\ref{ZComp}) we obtain
 \begin{equation}
 \mathcal{U}_W\left(\mathbf{x},k\right) = e^{i\hat{\phi}^a \left(\mathbf{x},k\right)\psi^{\dagger}_m\left(\mathbf{x}\right)T^a_{mn}\psi_n\left(\mathbf{x}\right)}=
 e^{i\hat{\phi}^a Q^a},
 \end{equation}
an interaction of the ``vector potential" $\phi^a$ with the fermionic non-Abelian charge $Q^a$.

 Note that
 \begin{equation}
 \mathcal{U}_W\left(\mathbf{x},k\right) \psi^{\dagger}_n\left(\mathbf{x}\right) \mathcal{U}^{\dagger}_W\left(\mathbf{x},k\right) =
 \psi^{\dagger}_m\left(\mathbf{x}\right) U_{mn}\left(\mathbf{x},k\right).
 \end{equation}
 Thus, if we define the fermionic tunneling Hamiltonian
 \begin{equation}
 H_t\!\left(\mathbf{x},k\right) \!=\! \lambda_{GM} \left(\psi^{\dagger}_m\!\left(\mathbf{x}\right) \psi_m\!\big(\mathbf{x}+\hat{\mathbf{k}}\big) \!+\! H.c.\right)
 \end{equation}
 the tunneling Hamiltonian $H_{GM}$ may be obtained by
 \begin{equation}
 \label{EqHGM}
 H_{GM} \left(\mathbf{x},k\right) = \mathcal{U}_W\left(\mathbf{x},k\right) H_t\left(\mathbf{x},k\right) \mathcal{U}^{\dagger}_W\left(\mathbf{x},k\right).
 \end{equation}
 This relation is the key ingredient for achieving the gauge-matter interactions, which are three body interactions, by using two body local interactions.
 In fact, this is a ``gauging" transformation, mapping a free fermionic tunneling term into a charged interaction term, by using a gauge transformation whose parameter is an operator. Examples may be found below, where we discuss particular gauge groups.

 One way to realize $H_{GM} \left(\mathbf{x},k\right)$ is the following. First, make the link $(\mathbf{x},k)$ and the vertex $\mathbf{x}$ interact, to generate $\mathcal{U}^{\dagger}_W\left(\mathbf{x},k\right)$.
 In the next step, allow the fermions to tunnel along the link, with the Hamiltonian $H_t\left(\mathbf{x},k\right)$ for time $\tau$, i.e. act with the unitary evolution
 \begin{equation}
 \mathcal{U}_t\left(\mathbf{x},k\right) = e^{-i H_t \left(\mathbf{x},k\right) \tau}.
 \label{Utdef}
 \end{equation}
 Finally make the link and the vertex interact again, but this time generate $\mathcal{U}_W\left(\mathbf{x},k\right)$.

Once again, the question is how to realize $H_{GM}$ for the whole lattice. From \eq{EqHGM}, we can already see that we have to respect some restrictions. For example, when we consider a single link $(\mathbf{x},1)$, we only have to apply $\mathcal{U}_W\left(\mathbf{x},1\right)$ but not $\mathcal{U}_W\left(\mathbf{x}+\mathbf{1},1\right)$, otherwise we get an interaction also with the gauge field of link $(\mathbf{x}+\mathbf{1},1)$. Therefore, we cannot act with $\mathcal{U}_W\left(\mathbf{x},1\right)$ for all $\mathbf{x}$ simultaneously. For a similar reason, we cannot act with both $\mathcal{U}_W\left(\mathbf{x},1\right)$ and $\mathcal{U}_W\left(\mathbf{x},2\right)$ at the same time because $H_{GM} \left(\mathbf{x},1\right)$ would get a contribution from the gauge field of link $(\mathbf{x},2)$ and this is not what we want.

This implies that we must divide the dynamics to four parts at least - this is the most economical
 way, and this is how we shall do it. We consider four sets of links - even horizontal (eh), even vertical (ev), odd horizontal (oh) and odd vertical (ov), named after the parity of the link
 from which they emanate, and their direction. The four sets have to be evolved one after the other according to the recipe \eqref{EqHGM}, but for all links in a given set we can realize the gauge-matter interactions simultaneously. Note that in this way we split $H_{GM}$ into four non-commuting terms. This will be taken into account when we discuss the digitization in more details.

 Let us consider the eh case for example. This is created with the following sequence:
 \begin{enumerate}
 \item Move all the gauge degrees of freedom on the even horizontal links to the beginning of the link, where they interact with the fermions in
 a way that generates
 \begin{equation}
 \mathcal{U}^{eh\dagger}_W=\underset{\mathbf{x}\text{ even}}{\prod} \mathcal{U}^{\dagger}_W\left(\mathbf{x},1\right).
 \end{equation}
 Then bring them back to their ``rest position" on the link.
 \item Allow tunneling on these links, for time $t$,  realizing
 \begin{equation}
 \mathcal{U}_t^{eh} = \underset{\mathbf{x}\text{ even}}{\prod} \mathcal{U}_t \left(\mathbf{x},1\right).
 \end{equation}
 \item Move the gauge degrees of freedom to the beginning of the link again, to interact with the fermions, this time for $\mathcal{U}^{eh}_W$.
 \end{enumerate}
 The result of this sequence is
 \begin{equation}
 \mathcal{U}^{eh}_W \mathcal{U}_t^{eh} \mathcal{U}^{eh\dagger}_W = W_{eh} = e^{-i \underset{\mathbf{x}\text{ even}}{\sum}H_{GM}\left(\mathbf{x},1\right)\tau}.
 \end{equation}

 Similarly, one may obtain $W_{ev},W_{oh},W_{ov}$.

 There is also another possibility, which makes use of stators. One first creates a stator for the relevant link, then let it interact with the fermion to generate
 $\widetilde{\mathcal{U}}^{eh\dagger}_W$ instead of $\mathcal{U}^{eh\dagger}_W$ (i.e., it involves the control instead of the physical degree of freedom). Similarly, after the fermionic tunneling
 one has to realize $\widetilde{\mathcal{U}}^{eh}_W$. The final step is to undo the stator, and the result is $W_{eh}$ again:
 \begin{equation}
 \begin{aligned}
 &\mathcal{U}^{\dagger}_{eh} \widetilde{\mathcal{U}}^{eh}_W \mathcal{U}_t^{eh} \widetilde{\mathcal{U}}^{eh\dagger}_W \mathcal{U}_{ eh} \left|\psi\right\rangle |\widetilde{in}\rangle
 \!=\! \mathcal{U}^{\dagger}_{eh} \widetilde{\mathcal{U}}^{eh}_W \mathcal{U}_t^{eh} \widetilde{\mathcal{U}}^{eh\dagger}_W S \left|\psi\right\rangle \\
 &= \mathcal{U}^{\dagger}_{eh} S \mathcal{U}^{eh}_W \mathcal{U}_t^{eh} \mathcal{U}^{eh\dagger}_W \!=\! W_{eh} \left|\psi\right\rangle |\widetilde{in}\rangle
 \end{aligned}
 \end{equation}

\subsubsection{The local terms}
The remaining parts of the Hamiltonian are ${ W_E = e^{-i H_E \tau} }$ and ${ W_M = e^{-i H_M \tau} }$. These are local terms which involve no interaction, and thus can be implemented directly by
acting locally on the physical degrees of freedom, either simultaneously or separately since they commute.

\subsubsection{The complete sequence}
With the interactions described above, we can write down the complete $\tau$ time step,
\begin{equation}
\label{EqSeq}
W(\tau) = W_{M} W_{E} W_{ov} W_{oh} W_{ev} W_{eh} W_{B}
\end{equation}
where $W_B = W_{Be}W_{Bo}$ (note that $\left[W_{Be},W_{Bo}\right]=0)$. Note that $W_{ov}, W_{oh}, W_{ev}, W_{eh}$  do not commute with each another.
There are other non-commuting ingredients as well, such as $W_{E}$ with $W_B, W_{ov}, W_{oh}, W_{ev}, W_{eh}$, and $W_M$ with $W_{ov}, W_{oh}, W_{ev}, W_{eh}$.

\subsection{Examples - Abelian theories}

In the scheme described above, we have discussed the most general situation, including the case where the group $G$ is non-Abelian. Particular non-Abelian cases result directly from that. In this section we will give instead examples of lattice gauge theories with an Abelian gauge group $G$.

The first example we show is for the $U(1)$ case - compact QED, which is somewhat more intuitive.
 It involves infinite dimensional local Hilbert spaces, both for the links and the controls, since the gauge group is continuous \cite{KogutLattice}. For this reason, the quantum simulation of a $U(1)$ theory is not feasible in practice. On the other hand, its truncation ($\mathbb{Z}_N$) involves finite Hilbert spaces and can be simulated in an experiment. This will be the next example we discuss.

\subsubsection{$U(1)$ (compact QED)}
The Hilbert space for the gauge field on each link is that of a particle on a ring: it may be described by an angle $\hat{\phi}$ or an angular momentum operator with an integer spectrum, $L$,
\begin{equation}
L\left|m\right\rangle = m \left|m\right\rangle.
\end{equation}
These operators are canonically conjugate:
\begin{equation}
\left[\hat{\phi},L\right]=i
\end{equation}
and $\hat{\phi}$ is the vector potential, while $L$ is the electric field. Thus, $\left|m\right\rangle$ are electric flux states - as already deduced before for the general Lie group case, but in this case it is much more intuitive.

From representing $L$ in differential form, $L=-i\partial_{\phi}$, we find out that the wavefunctions, or the Fourier transform, are given by
\begin{equation}
\left\langle \phi | m \right\rangle = \frac{1}{\sqrt{2\pi}}e^{i m \phi}.
\end{equation}

Just like in the $\mathbb{Z}_N$ case, the $U$ operators are not matrices. It is straightforward to see that
\begin{equation}
U = e^{i \hat{\phi}}
\end{equation}
is a flux raising operator.

The staggered electric charge takes the form
\begin{equation}
Q\left(\mathbf{x}\right) = \psi^{\dagger}\left(\mathbf{x}\right)\psi\left(\mathbf{x}\right) - \frac{1}{2}\left(1-(-1)^{x_1+x_2}\right)
\end{equation}
and the Gauss law operator (generator of gauge transformation) is simply
\begin{align}
G\!\left(\mathbf{x}\right) \!= & L\left(\mathbf{x},1\right) \!+\! L\left(\mathbf{x},2\right) \nonumber \\
&-\! L\left(\mathbf{x} \!-\! \hat{1},1\right) \!-\! L\left(\mathbf{x} \!-\! \hat{2},2\right) \!-\! Q\left(\mathbf{x}\right).
\end{align}

The Hamiltonian ingredients, in this case, take the form
\begin{align}
&H_{E} \!=\! \lambda_{E}\underset{\mathbf{x},k}{\sum}L^2\left(\mathbf{x}\right) \nonumber \\
&H_{B} \!=\! \lambda_{B}\underset{\mathbf{x}}{\sum}
\cos\!\left(\!\hat{\phi}\!\left(\mathbf{x},1\right) \!+\! \hat{\phi}\!\left(\mathbf{x} \!+\! \hat{1},2\right) \!-\! \hat{\phi}\!\left(\mathbf{x} \!+\! \hat{2},1\right) \!-\! \hat{\phi}\!\left(\mathbf{x},2\right)\!\right) \nonumber \\
&H_{M} \!=\! M\underset{\mathbf{x}}{\sum}(-1)^{x_1+x_2}\psi^{\dagger}\left(\mathbf{x}\right)\psi\left(\mathbf{x}\right) \nonumber \\
&H_{GM} \!=\! \lambda_{GM}\underset{\mathbf{x},k}{\sum}\left(\psi^{\dagger}\left(\mathbf{x}\right)e^{i\hat{\phi}\left(\mathbf{x},k\right)}\psi\left(\mathbf{x}+\mathbf{\hat{k}}\right)\!+\!H.c.\right)
\end{align}

Next we turn to the digital scheme. Note that not only the links, but also the controls are described by an infinite dimensional Hilbert space.
The initial state of the control is
\begin{equation}
\left|\tilde{in}\right\rangle = \left|\phi=0\right\rangle.
\end{equation}
If we let it interact with a link $i$, and generate the unitary operation
\begin{equation}
\mathcal{U}_i = \int d\phi\left|\phi_i\right\rangle\left\langle\phi_i\right| \otimes e^{-i \phi \tilde{L}} = e^{-i \hat{\phi}_i \tilde{L}}
\end{equation}
we obtain the stator
\begin{equation}
S_i = \int d\phi\left|\phi_i\right\rangle\left\langle\phi_i\right| \otimes \left|\tilde{\hat{\phi}}\right\rangle,
\end{equation}
for which
\begin{equation}
\tilde{U}S_i = S_i U.
\end{equation}

The local operation on the control, giving rise to the plaquette interaction, is simply
\begin{equation}
\tilde{H}_B = \lambda_B \cos\left(\tilde{\phi}\right)
\end{equation}
and the interaction of the links with the fermions, for $H_{GM}$, is
\begin{equation}
\mathcal{U}_W\left(\mathbf{x},k\right) = e^{i \hat{\phi}\left(\mathbf{x},k\right) \psi^{\dagger}\left(\mathbf{x}\right)\psi\left(\mathbf{x}\right)}
\end{equation}
as can be seen from
\begin{align}
&e^{i \hat{\phi}\left(\mathbf{x},k\right) \psi^{\dagger}\left(\mathbf{x}\right)\psi\left(\mathbf{x}\right)}
\psi^{\dagger}\left(\mathbf{x}\right)\psi\left(\mathbf{x+\hat{k}}\right)
e^{-i \hat{\phi}\left(\mathbf{x},k\right) \psi^{\dagger}\left(\mathbf{x}\right)\psi\left(\mathbf{x}\right)}
=\nonumber\\&\psi^{\dagger}\left(\mathbf{x}\right)e^{i \hat{\phi}\left(\mathbf{x},k\right)}\psi\left(\mathbf{x+\hat{k}}\right)
\end{align}

\subsubsection{$\mathbb{Z}_N$}
Next, we consider the case of finite Abelian groups - $\mathbb{Z}_N$ \cite{Horn1979}. As we shall describe in the next section, for small values of $N$ these are the first candidates for an experimental realization. We describe now the scheme for a general $N$. Note that in the limit $N\rightarrow\infty$, $\mathbb{Z}_N$ converges to $U(1)$, and thus this may be seen as a truncation scheme for the previously described $U(1)$ case (more details on this are found in the next section, where we discuss the experimental realization of $\mathbb{Z}_N$ gauge theories).

As the group is Abelian, it is irrelevant to talk about a representation, or about different left and right transformations. Thus, the representation basis for the gauge field will take the simple form $\left|m\right\rangle$, with a single integer quantum number, and dimension $N$. It is
 also redundant to describe $U$ as a matrix, and thus it will be, in the Abelian case, simply an operator acting on the gauge field's Hilbert space, which simplifies the mathematics a lot.

In the local gauge field Hilbert spaces, on each link, we define two unitary operators, $P$ and $Q$, satisfying
\begin{equation}
\begin{aligned}
P^N=Q^N&=1,\\
PQP^{\dagger}&= e^{i \frac{2\pi}{N}}Q,\\
Q\left|m\right\rangle &= \left|m+1\right\rangle \text{  (cyclically)},\\
P\left|m\right\rangle &= e^{i \frac{2\pi}{N} m}\left|m\right\rangle.
\end{aligned}
\end{equation}

On each vertex, there is a single fermionic species, $\psi^{\dagger}$. Due to the staggering, the gauge transformation is
\begin{equation}
\begin{aligned}
\Theta\left(\mathbf{x}\right) &= \! P\!\left(\mathbf{x},1\right)P\!\left(\mathbf{x},2\right)
P^{\dagger}\!\left(\mathbf{x} \!-\! \hat{1},1\right)P^{\dagger}\!\left(\mathbf{x} \!-\! \hat{2},2\right) \times \\
&e^{-i \frac{2\pi}{N}\left(\psi^{\dagger}\left(\mathbf{x}\right)\psi\left(\mathbf{x}\right) -\frac{1}{2}\left(1- (-1)^{x_1+x_2}\right)\right)}.
\end{aligned}
\end{equation}

The Hamiltonian terms take the forms
\begin{align}
&H_{E}=\lambda_{E}\underset{\mathbf{x},k}{\sum}\left(1-P\left(\mathbf{x},k\right)-P^{\dagger}\left(\mathbf{x},k\right)\right) \nonumber\\
&H_{B} \!=\! \lambda_{B}\underset{\mathbf{x}}{\sum} Q\!\left(\mathbf{x},1\right)Q\!\left(\mathbf{x} \!+\! \hat{1},2\right)Q^{\dagger}\!\left(\mathbf{x} \!+\! \hat{2},1\right)Q^{\dagger}\!\left(\mathbf{x},2\right) \!+\! H.c.\nonumber\\
&H_{M}=M\underset{\mathbf{x}}{\sum}(-1)^{x_1+x_2}\psi^{\dagger}\left(\mathbf{x}\right)\psi\left(\mathbf{x}\right) \nonumber\\
&H_{GM}=\lambda_{GM}\underset{\mathbf{x},k}{\sum}\left(\psi^{\dagger}\!\left(\mathbf{x}\right)Q\!\left(\mathbf{x},k\right)\psi\!(\mathbf{x} \!+\! \mathbf{\hat{k}}) \!+\! H.c.\right)
\end{align}

Let us describe the form of the interactions and local operations required for the realization of the digital sequence described above.
The control system will have the initial state
\begin{equation}
|\widetilde{in}\rangle = \frac{1}{\sqrt{N}}\underset{m}{\sum}\left|\widetilde{m}\right\rangle
\end{equation}
which is an eigenstate of Q, $\tilde{Q}|\widetilde{in}\rangle=|\widetilde{in}\rangle$.
The stator for a link $i$ will take the form
\begin{equation}
\begin{aligned}
S_i=\mathcal{U}_i |\widetilde{in}\rangle &= \frac{1}{\sqrt{N}}\underset{m}{\sum}Q^m \otimes \left|\widetilde{m}\right\rangle\\
\tilde{Q}S_i&=S_iQ^{\dagger}_i
\end{aligned}
\end{equation}
with the interaction
\begin{equation}
\mathcal{U}_i = \underset{m}{\sum}Q^m \otimes \left|\widetilde{m}\right\rangle \left\langle\widetilde{m}\right|.
\end{equation}
For the plaquette interactions, we have
\begin{equation}
\tilde{H}_B = \lambda_B \left(\tilde{Q}+\tilde{Q}^{\dagger}\right)
\end{equation}
and, finally, the interaction with the fermions is
\begin{equation}
\mathcal{U}_W\left(\mathbf{x},k\right) = e^{\log (Q\left(\mathbf{x},k\right)) \psi^{\dagger}\left(\mathbf{x}\right)\psi\left(\mathbf{x}\right)}.
\end{equation}

\section{Application for digital quantum simulation - the $\mathbb{Z}_3$ case}\label{appz3}

After having described the general scheme for the construction of digital lattice gauge theories, we will now discuss of a particular case, $\mathbb{Z}_3$, which may be used for
the construction of feasible quantum simulators. Although the group is finite, we want to stress that this and the example discussed in  \cite{ZNsim} can be seen as an important first step toward the simulation of compact Lie groups using the stator construction, as these examples may be seen as truncations of a compact Lie group ($U(1)$).

Previous proposals for quantum simulation of compact Lie groups have considered truncations of the gauge field infinite Hilbert space done in the so-called \emph{representation basis}. This is problematic for the use of stators since the $U$ operators in this particular truncation are not unitary anymore.  Therefore this method doesn't allow the construction of \emph{group element stators}. When one truncates a gauge group for the purpose of building a quantum simulator with the scheme proposed in the previous sections, the truncation should be done in the group element basis \cite{Zohar2015,ZoharStators} instead. For example, an approximation of $U(1)$ lattice gauge theories should be done in terms of $\mathbb{Z}_N$ theories (which converge to $U(1)$ in the $N \rightarrow \infty$ limit \cite{Horn1979}). Thus we shall describe here, as an example, the $\mathbb{Z}_{3}$ case (refer to \cite{ZNsim} for a similar proposal for $\mathbb{Z}_2$).

As we discussed for general $\mathbb{Z}_N$ theories,
the Hamiltonian parts $H_{B,e}$, ${ H_{B,o} }$, ${ H_{GM,eh} }$, ${ H_{GM,ev} }$, ${ H_{GM,oh} }$ and ${ H_{GM,ov} }$
 can be obtained effectively if we have a ancillary system that can interact with both the link and the matter degrees of freedom.

 These ancillary systems initially reside in the middle of every second plaquette (the even ones) and each of them is a ``copy"
  of the link degrees of freedom. In this particular example it is characterized by a 3-dimensional Hilbert space
  $ \tilde{\mathcal{H}}(\mathbf{x}) $, spanned by the basis vectors ${ \ket{\tilde{m}(\mathbf{x})} }$, ${ \tilde{m}=-1,0,1 }$ (this particular choice of the labeling becomes natural in light of the following discussion).
  Two operators ${ \tilde{P}(\mathbf{x}) }$ and ${ \tilde{Q}(\mathbf{x}) }$ can be defined similarly to what we did before.

\subsubsection{The stator}
\label{Sec:Z3Stator}

The explicit form of the stator for ${ \mathbb{Z}_3 }$ is
\begin{equation}
	\label{Eq:Z3StatorQ}
	S_{Q,i} \equiv \mathcal{U}_i | \widetilde{in} \rangle = \frac{1}{\sqrt{3}} \sum_{m=-1}^1 Q_i^{m} \otimes \ket{\widetilde{m}},
\end{equation}
with
\begin{equation}
	\label{Eq:Z3StatorQIn}
	| \widetilde{in} \rangle  =\frac{1}{\sqrt{3}} \sum_{m=-1}^1 \ket{\widetilde{m}}
\end{equation}
and
\begin{equation}
	\label{Eq:Z3StatorQUqp}
	\mathcal{U}_i = e^{i \frac{3}{2\pi} \log Q_i \log\widetilde{P}}.
\end{equation}

If we denote by ${ V_D }$ a local unitary transformation that maps the $P$-basis into the $Q$-basis, i.e. ${ Q = V_D^\dag P V_D }$,
we can write (\ref{Eq:Z3StatorQUqp}) as ${ \mathcal{U}_i = V_D^\dag \mathcal{U}'_i V_D }$, where
\begin{equation}
	\label{Eq:Z3StatorQUpp}
	\mathcal{U}'_i = e^{i \frac{3}{2\pi} \log P_i \log\widetilde{P}}.
\end{equation}
An even simpler form can be obtained if we note that ${ \log P = \frac{2\pi}{3\sqrt{3}} \left(P-P^\dag \right) }$, and then
\begin{equation}
	\label{Eq:Z3StatorQUpp}
	\mathcal{U}'_i = e^{i \frac{2\pi}{9} (P_i-P_i^\dag)(\tilde{P}-\tilde{P}^\dag)} .
\end{equation}

The ${ S_{Q,i} }$ stator satisfies the eigenoperator relation
\begin{equation}
	\label{Eq:Z3StatorQRelation}
	\tilde{Q} S_{Q,i} =  S_{Q,i} Q^{\dagger},
\end{equation}
which will have a key role in the rest of the this work. We can also define another type of stator - a \emph{P stator},
 \begin{equation}
 S_{P,i} = \tilde{V}_D S_{Q,i}
 \end{equation}
 where $\tilde{V}_D$ is the control system analog of $V_D$. $S_{P,i}$ satisfies the eigenoperator relation
 \begin{equation}
 \tilde{P} S_{P,i} = S_{P,i} Q^{\dagger}.
 \end{equation}
 The P stator will be used for implementing the gauge-matter interactions.

\subsection{The simulating system}
\label{Sec:Z3Simulating}

Next we shall describe how to control ultracold atoms in optical lattices~\cite{Jaksch2005,Bloch2008,Lewenstein2012} to implement the digital simulation of the $\mathbb{Z}_3$ Hamiltonian.

\subsubsection{The atomic ingredients}
\label{Sec:Z3Atoms}

\begin{figure}
  \centering
  % Requires \usepackage{graphicx}
  \includegraphics[width=0.3\textwidth]{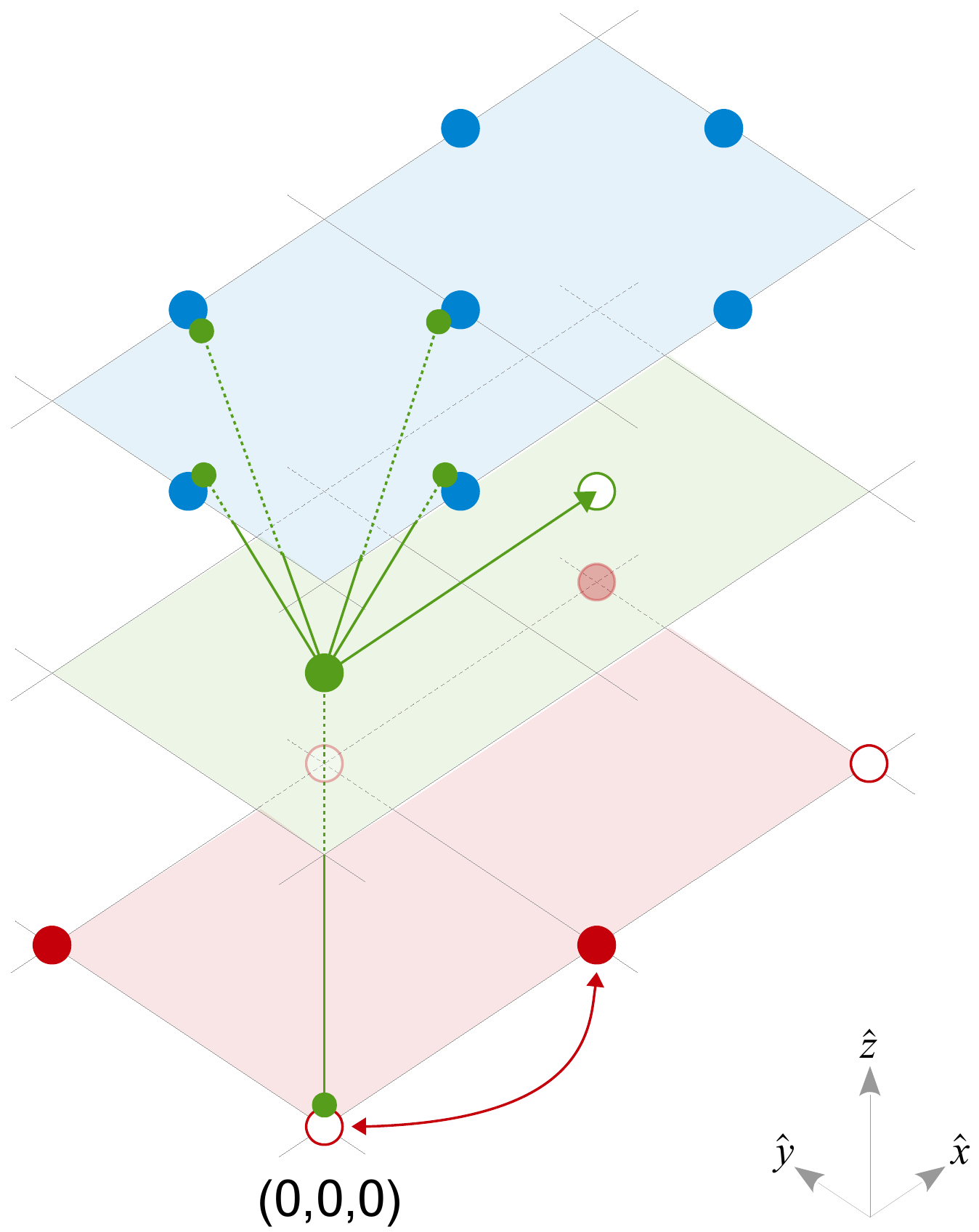}\\
  \caption{The $\mathbb{Z}_3$ simulating system consists of three layers, to avoid undesirable interactions. The lowest layer (red) is for the fermionic matter at the vertices,
  the highest one (blue) is for the link atoms (simulating the gauge field), and the ancilla atoms (green) are trapped "at rest" in an intermediate layer, from which they can rigidly move to interact with the other atoms, above or below them (see green lines). Full circles denote minima which are initially occupied. }\label{layersfig}
\end{figure}

To simulate the matter, it is natural to use fermionic atoms.
The trapping of the fermions in the desired positions can be achieved by creating a square optical lattice whose energy minima coincide with the sites of the simulated lattice.
The particular internal level of the fermions is not relevant, because the scheme described below does not involve any internal transition of the matter fermions.
Only the presence/absence of a fermion is an important degree of freedom. To describe creation matter fermions,
we use the fermionic operators ${ \psi^\dag\left(\mathbf{x}\right) }$.

To simulate the link (gauge field) degrees of freedom, we need an atomic species that allows us to control three of its internal level. For example, we propose to use alkali atoms
(bosons) that have an $F=1$ ground state, i.e. a threefold degenerate hyperfine level. These atoms can be trapped in the desired positions by creating a suitable optical
lattice whose minima coincide with the links of the simulated lattice. We anticipate that the link atoms and the matter fermions must be trapped
 on different vertical layers, to avoid undersired interactions with the moving ancilla atoms, which will "rest" in an intermediate layer. To describe creation of link atoms in a particular $m_F$ level, we use the operators ${ a_{m_F}^\dag\left(\mathbf{x},k\right) }$ where ${ m_F = -1,0,1 }$.

Finally, to simulate the auxiliary degrees of freedom, we need a third atomic species that  allows us to control three of its internal level as well.
For example, we propose to use another alkali species (still bosonic) that has an $F=1$ ground state, i.e. a threefold degenerate hyperfine level.
These atoms can be trapped in the desired positions by creating a suitable optical lattice whose minima coincide
with the center of every second plaquette of the simulated lattice. The control atoms must interact with both the matter fermions and the link atoms,
therefore they must be able to move between two different layers. To describe creation of auxiliary atoms in a particular $m_F$ level,
we use the operators ${ b_{m_F}^\dag\left(\mathbf{x},k\right) }$ where ${ m_F = -1,0,1 }$ again.

\begin{figure}
  \centering
  % Requires \usepackage{graphicx}
  \includegraphics[width=0.4\textwidth]{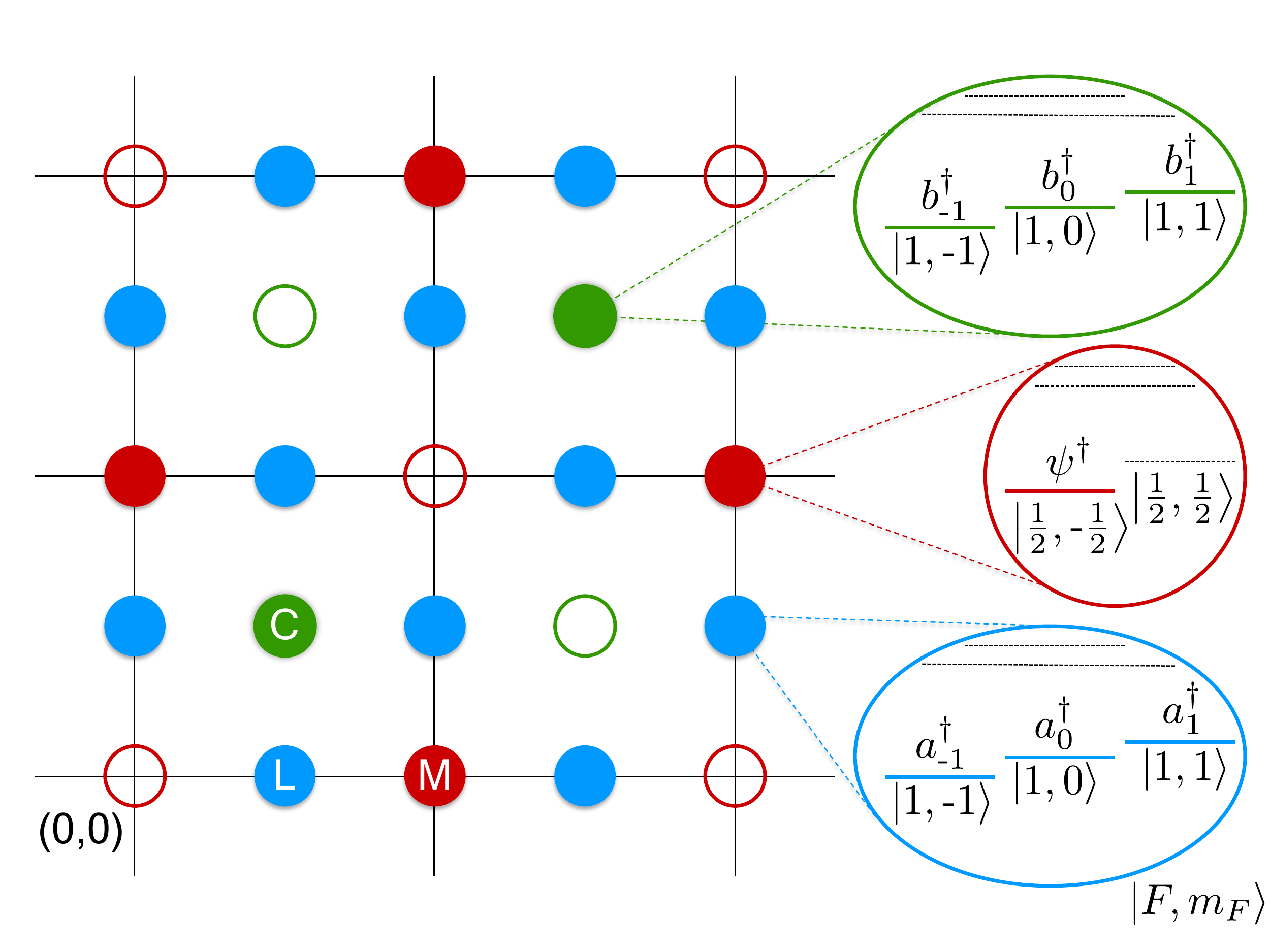}\\
  \caption{The $\mathbb{Z}_3$ simulating system consists of three atomic species, trapped in three different layers. The simulated system is planar, and contains gauge fields on the links (blue circles, with three possible atomic levels), matter fermions on the vertices (represented by red circles) and control, ancilla atoms (green) located at the centers of every second plaquette, as described in the text, with three internal levels as well.}\label{atomfig}
\end{figure}

The layer structure is shown in Fig. \ref{layersfig}. The corresponding atomic levels are presented in Fig. \ref{atomfig}.

\subsubsection{The optical lattices}
\label{Sec:Z3Lattices}

The atoms described above have to be trapped with the proper optical lattices, that we describe below.
All lengths are expressed in units of the simulated lattice spacing $s$, except where explicitly stated.

To trap the link atoms, we need the  optical potential
\begin{equation}
 V_{link}(x,y,z) = V_{link}(x,y) + V_{link}(z)
\end{equation}
 where ${ V_{link}(z) \propto (z - z_{link})^2 }$ provides harmonic vertical confinement around a single layer at position $z_{link}$ and
\begin{equation}
	\label{Eq:Z3PotentialLink}
	V_{link}(x,y) \propto \cos^2 \left( \pi (x+y) \right) +  \cos^2 \left( \pi (x-y) \right)
\end{equation}	
has minima for all positions ${ (x=m+\frac{1}{2},y=n) }$ and ${ (x=m,y=n+\frac{1}{2}) }$, i.e. in the middle of the links.
The optical lattice for the link atoms is therefore a square lattice, rotated by ${ 45^o }$ degrees with respect to the simulated lattice and with a lattice spacing equal to
$\frac{s}{\sqrt{2}}$. The minima have to be well separated in order to prevent tunneling and interactions between pairs of atoms in different positions.
This potential can be easily realized with two pairs of counter-propagating lasers (more details on the choice of the optical wavelengths are given in \app{App:Z3Lattices}).

To trap the matter fermions, we need the following optical potential ${ V_{mat}(x,y,z) = V_{mat}(x,y) + V_{mat}(z) }$ where ${ V_{mat}(z) \propto z^2 }$ provides harmonic vertical confinement around a single layer at position $z=0$ and
\begin{equation}
\begin{aligned}
	\label{Eq:Z3PotentialFermions}
	V_{mat}(x,y) & \! \propto \! \frac{1}{1+f(t)+h(t)} \cos^2 \left( \pi x +\frac{\pi}{2} \right) \\
	& +  \frac{1}{1+g(t)+h(t)} \cos^2 \left( \pi y +\frac{\pi}{2}\right) \nonumber \\
	& + \frac{f(t) + g(t) + h(t)}{1+ f(t)+ g(t) + h(t)} \cos^2 \left( \tfrac{\pi}{2} (x + y) + \phi(t) \right). \nonumber
\end{aligned}
\end{equation}

In the standard configuration, the fermions are trapped by well separated minima of equal depth, i.e. we have to set
 $f(t)=g(t)=h(t)=\phi(t)=0$ and prevent tunneling and interactions between pairs of atoms in different positions. The minima are then in all positions $(x=m,y=n)$
 and they coincide with the sites of the simulated lattice.
 Temporal shaping of the optical lattice through the functions $f(t), g(t), h(t), \phi(t)$ allows us to implement the Hamiltonians $H_M$ and $H_{GM}$,
  as we will discuss in the following (more details on the realization of $V_{mat}(x,y)$ are given in \app{App:Z3Lattices}).

To trap the auxiliary atoms, we need the optical potential
\begin{equation}
V_{aux}(x,y,z) = V_{aux}(x,y) + V_{aux}(z)
\end{equation}
 Here $V_{mat}(z) \propto (z-z_{aux}(t))^2$ provides harmonic vertical confinement around a single layer at position $z_{aux}(t)$, that can be varied in the range ${ [0,z_{link}] }$
 (remember that the auxiliary atoms have to interact with both the link and the matter atoms). In the other two dimensions, the potential is given by
\begin{align}
	\label{Eq:Z3PotentialAuxiliary}
	V_{aux}(x,y) & \propto \cos^2 \left( \frac{\pi}{2} x + \frac{\pi}{4} + \varphi_x(t) \right) \nonumber \\
	& +  \cos^2 \left( \frac{\pi}{2} y + \frac{\pi}{4} + \varphi_y(t) \right) \quad .
\end{align}
In the standard configuration, we have to set $\varphi_x(t)=\varphi_y(t)=0$ and the minima are in all positions ${ (x=2m+\frac{1}{2},y=2n+\frac{1}{2}) }$, i.e. in the middle of the even plaquettes of the simulated lattice. The minima have to be well separated in order to prevent tunneling and interactions
  between pairs of atoms in different positions. Temporal shaping of the optical lattice through the functions $\varphi_x(t),\varphi_y(t)$, $z_{aux}(t)$ allows us to move the auxiliary atoms to
  the middle of odd plaquettes, or to bring them close to the links and the matter atoms when interactions are needed (more details on the realization of $V_{aux}(x,y)$ are given
  in \app{App:Z3Lattices}).

\subsubsection{The interactions: atomic collisions}
\label{Sec:Z3Scattering}

When needed, interactions between different atoms can be implemented by bringing two atoms together and letting them scatter (collide).
In the case of ultracold atoms (${ T < 1 }$ mK), the
kinetic energy is small compared to any centrifugal barrier and there can only be s-wave scattering (total angular momentum ${ L=0 }$). Moreover, the kinetic energy
is typically smaller than the hyperfine splitting, so if two atoms are initially in their ground state, the transition probability to other hyperfine levels is strongly decreased.~\cite{Ho1998}.

Suppose that our two atoms have initial states ${ \ket{F_1,m_{F,1}} }$ and ${ \ket{F_2,m_{F,2}} }$ (we choose to denote with $m_F$ the projection of the
angular momentum along the $z$ direction, i.e. the different values of $F_z$). For symmetry reasons, the scattering potential commutes with the total angular momentum
${ \mathbf{F}_{tot}^2 = (\mathbf{F}_1 + \mathbf{F}_2)^2}$ and its projection on the $z$ direction
${ M_{F,tot} = m_{F,1} + m_{F,2} }$, which are then conserved quantities.
Moreover, we have seen that energetic constraints imply conservation of individual angular
 momenta ${ F_1^2 }$, ${ F_2^2 }$.
 Therefore, the allowed scattering processes are
 ${ \ket{F_1,m_{F,1}} \rightarrow \ket{F_1,m'_{F,1}} }$ and ${ \ket{F_2,m_{F,2}} \rightarrow \ket{F_2,m'_{F,2}} }$ with ${ m_{F,1} + m_{F,2} = M_{F,tot} = m'_{F,1} + m'_{F,2} }$.

The scattering potential explicitly depends on the total angular momentum $F_{tot}$, and can be approximated by the following
 pseudopotential~\cite{Ho1998,Ohmi1998,Stamper-Kurn2001}
\begin{equation}
	\label{Eq:Z3Scattering}
	V_{scat} \left( \mathbf{x} \right) = \frac{2 \pi }{\mu} \delta\left( \mathbf{x} \right) \sum_{F_{tot}} a_{F_{tot}} \mathbb{P}_{F_{tot}},
\end{equation}
where $\mathbf{x}$ is the relative position of the two atoms, $\mu$ is the reduced mass of the two colliding atoms,
 $a_{F_{tot}}$ is the scattering length for a particular value of ${F_{tot}}$ and $\mathbb{P}_{F_{tot}}$ is the projector onto the subspace corresponding to this particular value. The total angular momentum ${F_{tot}}$ can take all positive integer values between ${ \left| F_1 - F_2 \right| }$ and ${ F_1 + F_2 }$.

As an example, we can consider the interaction between the link and auxiliary degrees of freedom, that is a key step in the generation of the stators. Therefore, we will identify $F_1$ with $F$ and $F_2$ with $\tilde{F}$. If we assume that the two (nonidentical) atoms have ${ F = \tilde{F} = 1 }$, we can express the pseudopotential in the alternative form
\begin{equation}
	\label{Eq:Z3ScatteringFF}
	V_{scat} \left( \mathbf{x} \right) = \frac{2 \pi }{\mu} \delta\left( \mathbf{x} \right) \sum_{j=0}^2 g_j \left( \vec{F} \cdot \vec{\tilde{F}} \right)^j,
\end{equation}
where ${ g_0 = \tfrac{1}{3} (a_2 + 3 a_1 - a_0) }$, ${ g_1 = \tfrac{1}{2} (a_2 - a_1) }$ and ${ g_2 = \tfrac{1}{6} (a_2 - 3 a_1 + 2 a_0) }$.

If the optical trapping is insensitive to the value of $m_F$ and all components of the spin move together, the overlap of the atomic
(Wannier) wavefunctions will also be independent of $m_F$ when we bring two atoms close to each other.
Denoting this overlap as $\mathcal{O}(t)$ (accounting for its temporal dependence), the scattering interaction can be finally written as
\begin{equation}
	\label{Eq:Z3ScatteringUnitary}
	\mathcal{U}_{scat} = e^{-i \alpha \sum_{j=0}^2 g_j \left( \vec{F} \cdot \vec{\tilde{F}}  \right)^j},
\end{equation}
where ${ \alpha = \frac{2 \pi }{\mu} \int_0^{T_0} \! \mathcal{O}(t) dt }$ is the overlap  integrated over the whole interaction duration $T_0$. We would like to make this unitary operator equal to the one in \eqref{Eq:Z3StatorQUpp},
 as required for the creation of the stator.
 To make the comparison easier, we rewrite (\ref{Eq:Z3StatorQUpp})
 by inserting the identity
 \begin{equation}
 F_{z}=-\frac{i}{\sqrt{3}}\left( P-P^\dag \right)
 \end{equation}
 and obtain
\begin{equation}
	\label{Eq:Z3StatorFF}
	\mathcal{U}' = e^{-i \frac{2\pi}{3} F_z \tilde{F}_z} .
\end{equation}

By inspection of Eq. \eqref{Eq:Z3ScatteringFF},
 we see that there are many more terms.
 For example, we have ${ F_x \tilde{F}_x }$ and ${ F_y \tilde{F}_y }$ that involve changes in the values of $m_{F}$ and $\tilde{m}_{F}$. To suppress these processes,
 we can give them an energy penalty.
 This can be done, for example, by adding an energy term ${ E(m_F) = \mathcal{E} m_F }$ to the Hamiltonian of the link atoms,
  and similarly ${ \tilde{E}(\tilde{m}_F) = \tilde{\mathcal{E}} \tilde{m}_F }$ to the Hamiltonian of the auxiliary atoms (this can be achieved through a Zeeman splitting -
  more details will be given in the next section).
  Since the atomic collision conserves the quantity ${ M_F = m_F + \tilde{m}_F }$, the transition ${ m_F \rightarrow m'_F }$ is
  characterized by an energy cost ${ (m'_F - m_F)(\mathcal{E} - \tilde{\mathcal{E}}) }$. By making ${ \mathcal{E} \neq \tilde{\mathcal{E}} }$,
   the only energetically allowed processes (within a rotating wave approximation) conserve both $m_{F}$ and $\tilde{m}_{F}$.
   If we constrain Eq. (\ref{Eq:Z3ScatteringUnitary}) to be diagonal in the ${ m_F, \tilde{m}_F }$ basis and express it in second quantization, we get

\begin{equation}
	\label{Eq:Z3ScatteringUnitaryDiag}
	\mathcal{U}_{scat} = e^{-i \alpha (\eta_0 N_{tot} \tilde{N}_{tot} + \eta_1 F_z \tilde{F}_z + \eta_2 N_0 \tilde{N}_0) },
\end{equation}

where ${ \eta_{0}=g_{0}+\frac{3}{2} g_{2}  }$, ${ \eta_{1}=g_{1}-\frac{1}{2} g_{2} }$, ${ \eta_{2}=3 g_{2} }$, ${ N_{tot} = \sum_{M_F} a^\dag_{m_F} a_{m_F} }$ and ${ N_{0} = a^\dag_{0} a_{0} }$. Since we have one atom per well and no tunneling is allowed, we can set ${ N_{tot}=\tilde{N}_{tot}=1 }$. Finally, if we choose ${ \alpha = \frac{2\pi}{3 \eta_1} }$ we get

\begin{equation}
	\label{Eq:Z3ScatteringUnitaryDiag2}
	\mathcal{U}_{scat} = e^{-i \frac{2\pi}{3} (\frac{\eta_0}{\eta_1} + F_z \tilde{F}_z + \frac{\eta_2}{\eta_1} N_0 \tilde{N}_0) }.
\end{equation}

The first term in the exponential gives rise to  a global phase and hence is not important.
The second term gives the desired interaction, as in \eq{Eq:Z3StatorFF}.
The third term introduces an extra phase when both atoms are in their ${ m_F=\tilde{m}_F=0 }$ level. This term is undesired and needs to be eliminated.

To achieve this, we can spatially separate the different $m_F$ components of the atoms vertically (e.g. through a magnetic field gradient, see next section for details),
in a way that guarantees that when we move the auxiliary atoms onto the link, only the ${ m_F=\tilde{m}_F=0 }$ components will overlap. This gives rise to a unitary evolution of the form
\begin{equation}
	\label{Eq:Z3ScatteringUnitary00}
	\mathcal{V}_{scat} = e^{-i \beta N_0 \tilde{N}_0 }.
\end{equation}

If we tune the overlap and the interaction time such that $\beta = 2\pi ( \kappa - \frac{ \eta_2}{3 \eta_1}) > 0$ ($\kappa \in \mathbb{Z}$ can be chosen as the smallest allowed integer)
 and combine \eq{Eq:Z3ScatteringUnitaryDiag2} with \eq{Eq:Z3ScatteringUnitary00} we finally get \eq{Eq:Z3StatorFF} (up to a global phase).
\begin{equation}
	\label{Eq:Z3ScatteringUnitaryFinal}
	\mathcal{V}_{scat} \mathcal{U}_{scat} = e^{-i \frac{2\pi}{3} (\frac{\eta_0}{\eta_1} + F_z \tilde{F}_z)} e^{-i 2 \pi \kappa N_0 \tilde{N}_0 }.
\end{equation}
Note that the last piece of \eq{Eq:Z3ScatteringUnitaryFinal} has no effect whatsoever.
 To undo the stator, we need the inverse action ${ \mathcal{U}'^\dag }$.
  This can be implemented by flipping locally the ${ \tilde{m}_F=1 }$ and the ${ \tilde{m}_F=-1 }$ levels of the auxiliary atoms, which is equivalent to mapping ${ \tilde{F}_z }$ into ${ -\tilde{F}_z }$.
  We denote the spin flipping by $\tilde{V}_F$, and it is achievable by addressing the control atoms locally with lasers or RF light. Then,
  \begin{equation}
  \tilde{V}_F \mathcal{U}_{scat} \tilde{V}_F^{\dagger} = \mathcal{W}_{scat}
  \end{equation}
  where
\begin{equation}
	\label{Eq:Z3ScatteringUnitaryDiag3}
	\mathcal{W}_{scat} = e^{-i \frac{2\pi}{3} (\frac{\eta_0}{\eta_1} - F_z \tilde{F}_z + \frac{\eta_2}{\eta_1} N_0 \tilde{N}_0) },
\end{equation}
and putting it together with \eq{Eq:Z3ScatteringUnitary00} again we get
\begin{equation}
	\label{Eq:Z3ScatteringUnitaryFinal2}
	\mathcal{V}_{scat} \mathcal{W}_{scat} = e^{-i \frac{2\pi}{3} (\frac{\eta_0}{\eta_1} - F_z \tilde{F}_z)} e^{-i 2 \pi \kappa N_0 \tilde{N}_0 }.
\end{equation}
This is exactly ${ \mathcal{U}'^\dag }$ up to a global phase.

\subsubsection{Magnetic fields}
\label{Sec:Z3Magnetic}

To lift the degeneracy of the ground state, we can use, for example, a uniform magnetic field $\mathbf{B}$, which gives rise to the magnetic perturbation
\begin{equation}
H_{Z} = \mu_{B} \mathfrak{g}_{F} m_{F} B
\end{equation}
 where $ \mu_{B} $ is the Bohr magneton and $\mathfrak{g}_{F}$ is the hyperfine Land\`e factor.
 However, a static magnetic field will split the levels of all atoms,
 so we must choose three atomic species that have different Land\`e factors. Another possibility is to split the levels
 with alternative methods that allow the addressing of a single species (e.g. the AC stark effect).

We can achieve even more control if we use a magnetic field gradient, for example $\mathbf{B}(z) = b z\hat{\mathbf{z}}$.
 In this case, different ${ m_{F} }$ levels will experience different vertical forces and as a consequence they will be localized around different vertical
 equilibrium positions within the harmonic well. This allows us to tailor the atomic collisions (in particular control how much the wavefunctions of different atomic components
 overlap with each other) and make them depend on the specific values of ${ m_{F,1} }$ and ${ \tilde{m}_F }$. If the separation between the different components is big enough,
 we can for example bring together only the ${ m_F=\tilde{m}_F=0 }$ levels and implement the ${ \mathcal{V}_{scat} }$ evolution described above (see \eq{Eq:Z3ScatteringUnitary00}).

A magnetic field gradient has been similarly employed in previous experiments to selectively trap only the ${ m_F=0 }$ component (which is insensitve to the magnetic field) while
 pushing all other components out of the trap~\cite{Zielonkowski1998a}, so a spatial splitting of the ${ m_F }$ components should be achievable with weaker gradients.

\subsection{Implementation of the digital dynamics}
\label{Sec:Z3Hamiltonian}

Having all the ingredients and techniques at hand, we can finally discuss the
implementation of the different parts of the $\mathbb{Z}_3$ Hamiltonian.

\subsubsection{Standard configuration of the lattice}
\label{Sec:Z3HamiltonianStandard}

First, let us describe the standard configuration of the lattice. The potentials ${ V_{link}(x,y,z) }$ and ${ V_{aux}(x,y,z) }$ are always active to trap the link and auxiliary
atoms in the desired positions. These  must be loaded with one atom per minimum.
 Tunneling and interactions are prevented by creating deep and well separated minima. Cooling of the atoms makes sure that they are all in their motional ground state with energy
 ${ E_{0,link} }$ and ${ E_{0,aux} }$ respectively. A uniform magnetic field ${ B_1 }$ (or an \mbox{AC-stark} effect) is present as well to lift the degeneracy of the ground state
 and induce energy splittings (${ \Delta E_{link} }$ and ${ \Delta E_{aux} }$ respectively) between the different ${ m_F, \tilde{m}_F }$ components.

 The control atoms are prepared in the state $\left|\tilde{in}\right\rangle$  of \eq{Eq:Z3StatorQIn}. The gauge field atoms on the links are prepared in the state $\left|0\right\rangle$ for
 their part in the global singlet state of \eq{globsing}.

 The potential ${ V_{mat}(x,y,z) }$ is in its standard configuration (see \sect{Sec:Z3Lattices}) and is half filled, with exactly one fermion in each odd minimum (energy ${ E_{0,mat} }$).
 This implements the Dirac state $\left|D\right\rangle$, and completes the global singlet state of \eq{globsing}.

This gives rise to the non-interacting Hamiltonian $H_0$, expressed here in second quantization
\begin{align}
	\label{Eq:Z3HamiltonianStandard}
	H_0 & = E_{0,mat} \sum_{\mathbf{x}} \psi^\dag(\mathbf{x}) \psi(\mathbf{x}) \\
	& + \sum_{\mathbf{x},k} \left( E_{0,link} \plus \Delta E_{link} m_F \right) a_{m_F}^\dag(\mathbf{x},k) a_{m_F}(\mathbf{x},k) \nonumber \\
	& + \sum_{\mathbf{x},k} \left( E_{0,aux} \plus \Delta E_{aux} \tilde{m}_F \right) b_{m_F}^\dag(\mathbf{x}) b_{m_F}(\mathbf{x}). \nonumber
\end{align}
Whenever we implement a piece of the $\mathbb{Z}_3$ Hamiltonian, this will be added to $H_0$ (which is always acting on the system).
To recover the desired Hamiltonian $H$ we must therefore move to an interaction picture and cancel the ``free evolution" $H_0$
- i.e., we will work in a rotating frame with respect to $H_0$ and make use of the rotating wave approximation.

\subsubsection{The electric Hamiltonian}
\label{Sec:Z3HamiltonianElectric}

The electric Hamiltonian ${ H_E }$, expressed in second quantization, reads
\begin{align}
	\label{Eq:Z3HamiltonianElectric}
	H_E & = \lambda_E \sum_{\mathbf{x},k} \left( 1 + \left| m_F \right| \right) a_{m_F}^\dag({\mathbf{x},k}) a_{m_F}({\mathbf{x},k}).
\end{align}
When acting with this Hamiltonian  for a short time $\tau$, as prescribed by the digitization, the evolution of the system is described by the operator
\begin{align}
	\label{Eq:Z3UnitaryElectric}
	W_E = e^{- i H_E \tau}.
\end{align}

Two terms can be easily identified in \eq{Eq:Z3HamiltonianElectric}. The first contains $\sum_{\mathbf{x},k} a_{m_F}^\dag({\mathbf{x},k}) a_{m_F}(\mathbf{x},k)$, which is a constant of motion. Therefore, it gives rise to a global phase and can be neglected. The second part, $\sum_{\mathbf{x},k} \left| m_F \right| a_{m_F}^\dag({\mathbf{x},k}) a_{m_F}(\mathbf{x},k)$ is the relevant piece that has to be implemented. Therefore we need to give an additional energy shift ${ \Delta E_{el} }$ to the ${ m_F = \pm 1 }$ levels of the links (the same shift for both levels), for example by the use of external lasers addressing these levels. This has to be done for  a time  ${ \frac{\lambda_E}{\Delta E_{el}} \tau}$.

\subsubsection{The mass Hamiltonian}
\label{Sec:Z3HamiltonianMass}

The mass Hamiltonian ${ H_M }$, expressed in second quantisation, reads
\begin{align}
	\label{Eq:Z3HamiltonianMass}
	H_M & = M \sum_{\mathbf{x}} (-1)^{m+n} \psi^\dag(\mathbf{x}) \psi(\mathbf{x}),
\end{align}
and the unitary evolution that we want to implement is
\begin{align}
	\label{Eq:Z3UnitaryMass}
	W_M = e^{- i H_M \tau}.
\end{align}

In our cold-atomic system, this can be achieved by shaping the optical lattice ${ V_{mat}(x,y,z) }$
through the function $h(t)$ (see \sect{Sec:Z3Lattices} and \app{App:Z3Lattices}), that must be tuned from the value $0$ to some
 value $h_0$. In this way, the energy of the even minima is raised by an amount $M_{even}$ and we can realize the Hamiltonian
\begin{align}
	\label{Eq:Z3HamiltonianMassImplemented}
	H'_M & = M_{even} \sum_{\mathbf{x}} (1+(-1)^{m+n}) \psi^\dag(\mathbf{x}) \psi(\mathbf{x}).
\end{align}
If we act with this Hamiltonian for a time ${ \frac{M_{even}}{M} \tau }$ we can obtain the required unitary evolution, up to a global phase that is not important.

\subsubsection{The even plaquette interactions}
\label{Sec:Z3HamiltonianPlaquette}

For the unitary evolution $W_{Be}$ (even plaquette interactions),
we first create a stator for the even plaquette - i.e., the unitary operation $\mathcal{U}_{pe}$ (\ref{Upedef}). Its ingredients $\mathcal{U}_{ie}$ (\ref{U1edef})-(\ref{U4edef})
may be obtained as
\begin{equation}
\mathcal{U}_i = V^\dag_{D,all} \prod_{\mathbf{x}\text{ even}} \mathcal{U}'_i\left(\mathbf{x}\right)  V_{D,all}
\end{equation}
where
\begin{equation}
	\label{Eq:Z3StatorQUqp2}
	\mathcal{U}'_i\left(\mathbf{x}\right) = e^{i \frac{3}{2\pi} \log P_i(\left(\mathbf{x}\right)) \log\widetilde{P}\left(\mathbf{x}\right) }.
\end{equation}
To realize the product of $\mathcal{U}_i$ operators,
 we rigidly shift the optical lattice of the auxiliary atoms so that each auxiliary atom gets very
 close to one of the link atoms. This happens in parallel for all even plaquettes, and only for even plaquettes because of the initial positions of the control atoms.
 The interaction process between two atoms has been extensively discussed in \sect{Sec:Z3Scattering} and can be controlled to give the desired evolution.
  We then repeat this process for all four links around a plaquette, with the right orientation, and obtain
\begin{equation}
\mathcal{U}_{pe}=V^{\dagger}_{D,all} \mathcal{U}'^{\dagger}_{4e} \mathcal{U}'^{\dagger}_{3e} \mathcal{U}'_{2e} \mathcal{U}'_{1e} V_{D,all}.
\end{equation}
${ V_{D,all} }$ is the unitary transformation that changes from a $P$-basis to a $Q$-basis for all link atoms.
We can express it in the alternative form ${ V_{D,all} = e^{-i H_D \frac{\pi}{2 \sqrt{3}}} }$, where

\begin{align}
	\label{Eq:Z3VD}
	H_{D} = \sum_{\mathbf{x},k} \Big[ & \left(1-\sqrt{3}\right) a_0^\dag(\mathbf{x},k) a_0(\mathbf{x},k)  \nonumber \\
	& -\frac{1}{2}\left(1+2\sqrt{3}\right) \left(  a_{\pm1}^\dag(\mathbf{x},k) a_{\pm 1}(\mathbf{x},k) \right) \\
	& + \Big( a_{1}^{\dagger}(\mathbf{x},k)a_{0}(\mathbf{x},k)+a_{-1}(\mathbf{x},k)^{\dagger}a_{0}(\mathbf{x},k) \nonumber \\
    &-\tfrac{1}{2}a_{1}(\mathbf{x},k)^{\dagger}a_{-1}(\mathbf{x},k) + H.c. \Big) \Big] \nonumber
\end{align}
is a local Hamiltonian that can be created by means of optical/radiofrequency fields.
Acting on the control's initial state with $\mathcal{U}_{pe}$, when the controls are in the centers of the even plaquettes,
 produces a plaquette stator $S_{Q,\square}.$
If we now evolve locally the control atoms with the unitary
\begin{equation}
	\widetilde{V}_{B}=e^{-i \tau \lambda_B\sum_{\mathbf{x}\text{ even}} \left(\widetilde{Q}(\mathbf{x})+\widetilde{Q}^{\dagger}(\mathbf{x})\right)}
\end{equation}
we obtain, using the stator, the even plaquette interactions, $W_{Be}$. For disentangling the stators, we can use $\mathcal{U}^{\dagger}_{pe}$.

To realize the odd plaquettes evolution ($W_{Bo}$), we can move the auxiliary atoms to the centers of odd plaquettes and repeat all of the above. Then we bring the auxiliary atoms back to the centers of the even plaquettes.

\subsubsection{Gauge-matter interaction on even horizontal links}
\label{Sec:Z3HamiltonianGaugeMatterEH}

We also generate the gauge-matter interactions with stators. Let us demonstrate how to do that, for example, in the case of even horizontal links.
For that,
we rigidly move the auxiliary
atoms to interact with the even horizontal link atoms and create a P-stator.
 Next, we move the auxiliary atoms to interact with the fermions in the beginning of the link. This interaction
is similar to the one described in \sect{Sec:Z3Scattering} (again an atomic collision) but this time we have $F_1 = F' = 1/2$ (the fermion) and $F_2 = \tilde{F} = 1$ (the auxiliary atom). The total angular momentum $F_{tot}$ can take only two values: $\left|\tilde{F}-F'\right|$ and $\tilde{F}+F'$.
This gives rise to a Hamiltonian of the form
\begin{align}
	H_{WW'} & = \mathcal{O}_{W}\left(t\right)\left(g'_{0}\psi^{\dagger}\psi \underset{m}{\sum}b_m^{\dagger}b_m +g'_{1}\psi^{\dagger}\psi\widetilde{F}^{z}\right) \nonumber \\
	& = \mathcal{O}_{W}\left(t\right)\left(g'_{0}\psi^{\dagger}\psi+g'_{1}\psi^{\dagger}\psi\widetilde{F}^{z}\right).
\end{align}
Once again we can say that $\underset{m}{\sum}b_m^{\dagger}b_m = 1$, which gives rise to the second equality. However, now the first term does not give rise to a global phase,
since the number of $\psi$ fermions is only a \emph{globally} conserved quantity. We use the relation
\begin{equation}
\widetilde{F}^{z}=-\frac{3i}{2\pi}\log\widetilde{P},
\end{equation}
to obtain
\begin{equation}
H_{WW'}=\mathcal{O}_{W}\left(t\right)\left(g'_{0}\psi^{\dagger}\psi-\frac{3i}{2\pi}g'_{1}\psi^{\dagger}\psi\log\widetilde{P}\right)
\end{equation}
and split it into
\begin{equation}
H_{W}=-\frac{3i}{2\pi}\mathcal{O}_{W}\left(t\right)g'_{1}\psi^{\dagger}\psi\log\widetilde{P}
\end{equation}
and
\begin{equation}
H_{W'}=\mathcal{O}_{W}\left(t\right)g'_{0}\psi^{\dagger}\psi.
\end{equation}
Then, if the interaction takes time $T_W$, and
\begin{equation}
\int_{0}^{T_{W}}\mathcal{O}_{W}\left(t\right)dt=-\frac{2\pi}{3g'_{1}}
\end{equation}
we get that $H_{W}$ gives rise to the unitary
\begin{equation}
\label{Eq:Z3AuxiliaryFermionInteraction}
\mathcal{\tilde{U}}^{\dagger}_W = e^{\psi^{\dagger}\psi\log\left(\widetilde{P}\right)}
\end{equation}
while $H_{W'}$ (which commutes with it) is responsible for
\begin{equation}
V_{W'}\left(\theta\right)=e^{-i\theta\psi^{\dagger}\psi}
\end{equation}
where
\begin{equation}
\theta=-\frac{8\pi}{2g'_{1}}g'_{0}.
\end{equation}

Note the influence of the $\tilde{F}_z$ flipping operation on the unitary operations that we have just introduced:
\begin{equation}
\begin{aligned}
\tilde{V}_F \mathcal{\tilde{U}}^{\dagger}_W \tilde{V}^{\dagger}_F &=\mathcal{\tilde{U}}_W \\
\tilde{V}_F V_{W'}\left(\theta\right) \tilde{V}^{\dagger}_F &= V_{W'}\left(\theta\right).
\end{aligned}
\end{equation}
With all these we can finally derive the gauge-matter interaction on the link. We begin with a P stator of the relevant link, $S_{Pi}$,
then let the control interact with the fermion at the beginning of the link in order to get $\mathcal{\tilde{U}}^{\dagger}_W$ and $V_{W'}\left(\theta\right)$. Next, we allow tunneling on the link and realize $\mathcal{U}_t$ (as defined in \eq{Utdef}). Finally, we flip the $\tilde{m}_{\tilde{F}}$ levels of the control, let it interact with the fermion again and re-flip its levels, obtaining $\mathcal{\tilde{U}}_W$ and $\mathcal{\tilde{U}}^{\dagger}_W$.
Altogether, acting on the P-stator, we have:
\begin{equation}
V_{W'} \! \left(\theta\right) \mathcal{\tilde{U}}_W \mathcal{U}_t \mathcal{\tilde{U}}^{\dagger}_W V_{W'}\!\left(\theta\right) S_P \!=\!
 S_P V_{W'}\!\left(\theta\right) \mathcal{U}_W \mathcal{U}_t \mathcal{U}^{\dagger}_W V_{W'}\!\left(\theta\right)
\end{equation}
The sequence $\mathcal{U}_W \mathcal{U}_t \mathcal{U}^{\dagger}_W$ will now give rise to the desired time evolution of $H_{GM}$ on the link. We are left, however, with some
fermion-dependent phases $V_{W'}\left(\theta\right)$, which, as we show in \app{App:B}, do not influence the physics and may be ignored once the entire lattice is considered.

Finally, we have to undo the P-stator if we want to disentangle the auxiliary and the link atoms and move to the next step.

\begin{figure*}
  \centering
  % Requires \usepackage{graphicx}
  \includegraphics[width=\textwidth]{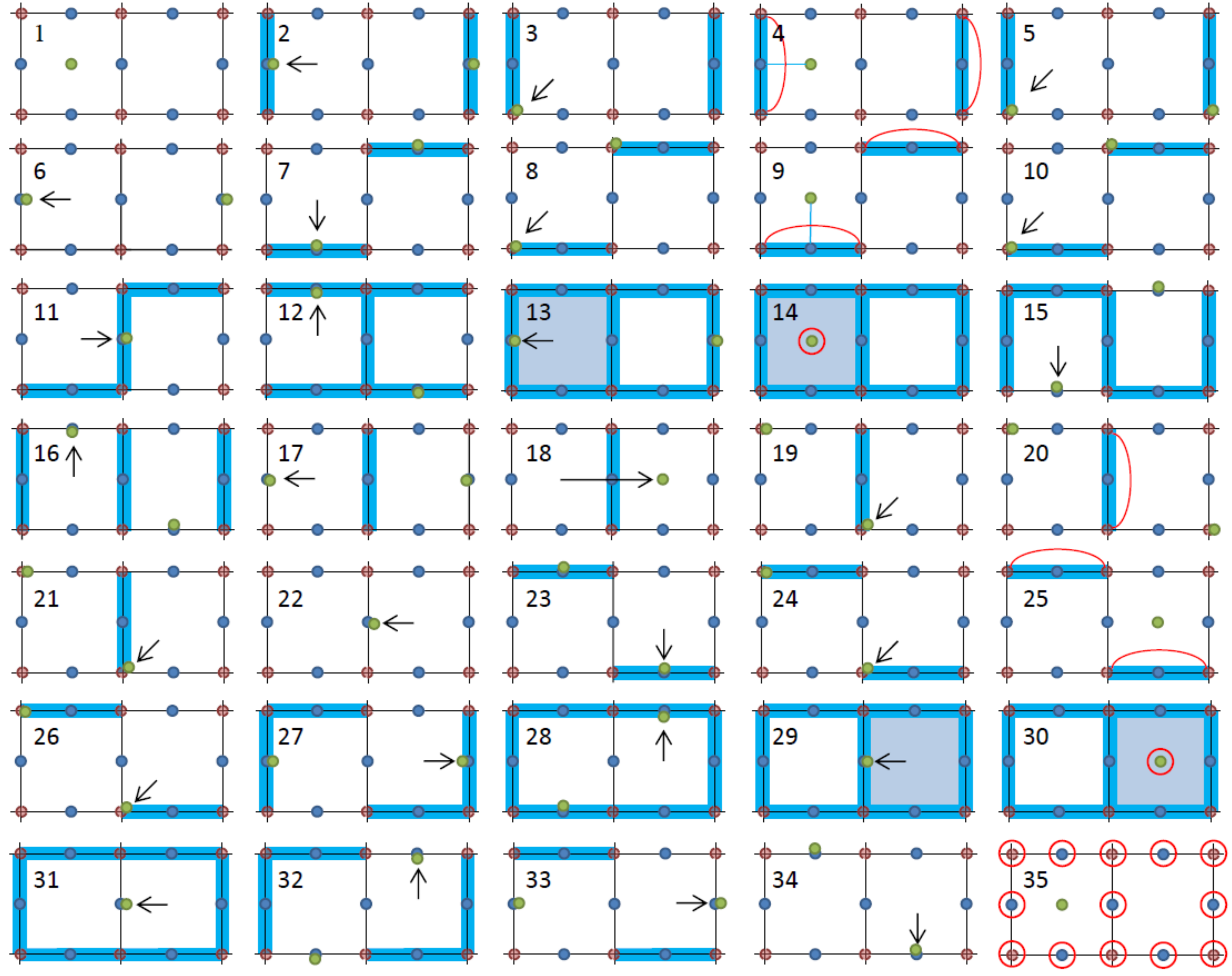}\\
  \caption{The complete $\mathbb{Z}_3$ sequence, for a single time-step $W$ (simulating time $\tau$). (1) Two plaquettes - the left is even, the right is odd. Blue circles denote the gauge field atoms, red ones the matter fermions (which can be present or absent) and green ones the control atoms, initially in the centers of the even plaquettes.
  (2) $\mathcal{U}_{ev}$, creating a stator for the even vertical links (here and in the following, blue lines/squares show that a stator is active for the corresponding links/plaquettes).
   \mbox{(3) The} controls interact with fermions to realize $\mathcal{U}^{ev\dagger}_W$.
   (4) Fermionic tunneling for even vertical links is allowed: $\mathcal{U}_t$ (here and in the following denoted by a red line connecting a pair of fermions).
   \mbox{(5) $\mathcal{U}^{ev}_W$.}
   \mbox{(6) The stator is undone.}
   (7)-(10) A similar process for even horizontal links, but without undoing the stators.
   \mbox{(11)-(13) Plaquette stators are completed for even plaquettes.}
   (14) $\tilde{V}_B$ is generated by addressing the control atoms (local operations on atoms are denoted by red circles)}. This will eventually give rise to $W_{Be}$.
   (15)-(17) The plaquettes' stators are undone, but stators for odd vertical links are kept.
   \mbox{(18) The controls' ``rest positions" are moved to the odd plaquettes on the right.
   (19-34) Similar process is repeated for the odd plaquettes, but the stators eventually are completely undone in this case, for the next step.
   (35) The controls are moved back to the centers of the even plaquettes, and the link and vertex atoms are evolved according to the non-interacting parts, $W_E$ and $W_M$ (again non-interacting terms are denoted by red circles). Then the whole sequence can be repeated.
   }
  \label{figseq}
\end{figure*}

\subsubsection{The complete sequence}
We can now summarize and give the complete sequence of operations required for a single time step $W$ of the $\mathbb{Z}_3$ simulation:
\begin{enumerate}
  \item Start with the control atoms in the centers of even plaquettes, $\mathbf{x}$. First, we wish to obtain $W_{ev}$.
 For that, we do the following:
        \begin{enumerate}
        \item Act with the sequence
        $\mathcal{U}_{ev} = \tilde{V}_D V_D^{\dagger} \mathcal{U}'_{ev} V_D$,
        to create P-stators for the even vertical links (which are on the left of the control atoms).
        \item Move the controls to interact with the fermions on the beginning of the even vertical links - i.e., obtain
        $\mathcal{U}^{ev\dagger}_{W} e^{-i \theta \underset{\mathbf{x}\text{ even}}{\sum}\psi^{\dagger}\left(\mathbf{x}\right)\psi\left(\mathbf{x}\right)}$.
        \item Perform $\mathcal{U}_t^{ev}$.
        \item Flip the control's $\tilde{m}_F$ with $\tilde{V}_F^{\dagger}$, then interact with the fermions again and flip again, for
        $\mathcal{U}^{ev}_{W} e^{-i \theta \underset{\mathbf{x}\text{ even}}{\sum}\psi^{\dagger}\left(\mathbf{x}\right)\psi\left(\mathbf{x}\right)}$.
        \item Act with $\tilde{V}^\dag_D$ to convert the stator into a Q-stator.
        \item Act with $\mathcal{U}_{ev}^{\dagger}$ to undo the stator.
        \end{enumerate}
        The result is then
  \begin{equation}
        \left|\tilde{in}\right\rangle W'_{ev}
  \end{equation}
  where $W'_{ev}=e^{-i\theta\underset{\mathbf{x}even}{\sum}\psi^{\dagger}\left(\mathbf{x}\right)\psi\left(\mathbf{x}\right)}W_{ev}e^{-i\theta\underset{\mathbf{x}even}{\sum}\psi^{\dagger}\left(\mathbf{x}\right)\psi\left(\mathbf{x}\right)}$.
  \item Repeat $1(a)-1(e)$ for the link below the stator (even horizontal), to obtain
  \begin{equation}
  \mathcal{U}_{eh}\left|\tilde{in}\right\rangle W'_{eh}W'_{ev}
  \end{equation}
  \item Even plaquette interactions:
        \begin{enumerate}
        \item Complete the plaquette stators, acting with $\mathcal{U}_{ev}^{\dagger}\mathcal{U}_{oh}^{\dagger}\mathcal{U}_{ov}$.
        \item Act locally on the controls with $\tilde{V}_B$.
        \item Undo most of the plaquette's stator, acting with $\mathcal{U}_{ev}\mathcal{U}_{oh}\mathcal{U}_{eh}^{\dagger}$
        \end{enumerate}
  After step 3 we are left with:
  \begin{equation}
  \mathcal{U}_{ov}\left|\tilde{in}\right\rangle W_{Be} W'_{eh}W'_{ev}
  \end{equation}
  \item Move the controls to the right - to the centers of the odd plaquettes. To avoid undesired interactions between the link and the auxiliary degrees of freedom during the movement, it is important that the two atomic species live on different vertical layers.
  \item Create the odd vertical link interactions:
        \begin{enumerate}
        \item Act with $\tilde{V}_D$ to convert the stators into P stators.
        \item Repeat $1(b)-1(f)$ with the link on the left of the control.
        \end{enumerate}
  The result:
  \begin{equation}
  \left|\tilde{in}\right\rangle W'_{ov} W_{Be} W'_{eh}W'_{ev}
  \end{equation}
  \item Repeat $1(a)-1(e)$ with the link below the control, to obtain the interactions for odd horizontal links. The result is
  \begin{equation}
  \mathcal{U}_{oh}\left|\tilde{in}\right\rangle W'_{oh} W'_{ov} W_{Be} W'_{eh}W'_{ev}
  \end{equation}
  \item Odd plaquette interactions:
        \begin{enumerate}
        \item Complete the plaquette stators, acting with $\mathcal{U}_{ov}^{\dagger}\mathcal{U}_{eh}^{\dagger}\mathcal{U}_{ev}$.
        \item Act locally on the controls with $\tilde{V}_B$.
        \item Undo completely the plaquette stator, acting with $\mathcal{U}_{oh}^{\dagger}\mathcal{U}_{ev}^{\dagger}\mathcal{U}_{eh}\mathcal{U}_{ov}$.
        \end{enumerate}
  The result:
  \begin{equation}
  \left|\tilde{in}\right\rangle W_{Bo} W'_{oh} W'_{ov} W_{Be} W'_{eh}W'_{ev}
  \end{equation}
  \item Complete the sequence with the non interacting steps (mass and electric Hamiltonians), to obtain
  \begin{equation}
  \left|\tilde{in}\right\rangle W' \!=\!  \left|\tilde{in}\right\rangle W_E W_M W_{Bo} W'_{oh} W'_{ov} W_{Be} W'_{eh}W'_{ev}
  \label{Eq:Sequence}
  \end{equation}
\end{enumerate}
 The complete sequence is described in Fig. \ref{figseq}.

 Considering the commutation relations of some of the steps, discussed above, we finally obtain
\begin{equation}
W' = W_E W_M W_B W'_{oh} W'_{ov} W'_{eh}W'_{ev}
\end{equation}
and, within our physical Hilbert space, this is physically equivalent to
\begin{equation}
W(\tau) = W_E W_M W_B W_{oh} W_{ov} W_{eh}W_{ev}
\end{equation}
as shown in \app{App:B}. If we put $\tau=t/M$ and repeat the sequence $W(\tau)$ for $M$ consecutive times, we obtain the trotterized evolution described in section \ref{SecDigit}.

\subsection{Limitations}
\label{Sec:Z3Errors}

Several kinds of errors can affect the precision of the simulation. On one hand, we have intrinsic errors coming from the digitization. The sequence described above is only equal to the desired evolution to first approximation, and errors scaling as $t^2/M$ arise because the various pieces of the sequence do not commute. However, this error can be made as small as desired by increasing the number of steps $M$, at the cost of a longer simulation time. It's also important to note that each piece of the sequence, and hence their commutators, is gauge invariant so the digitized evolution of the simulating system respects in principle the desired symmetry.

On the other hand, there might be experimental imperfections in the implementation of the desired local evolutions and interactions. Importantly, such errors can break the symmetry and tend to accumulate for a large number of steps $M$. Therefore, care should be taken in minimizing their effect, especially for errors affecting gates that take a fixed amount of time, independent of $M$, such as $V_F$, $\mathcal{U}_{scat}$, $\mathcal{V}_{scat}$, etc. For the same reason, $M$ should not be increased arbitrarily but must be chosen to balance the digitization and the implementation errors (see discussion below).

\subsubsection{Trotterization}
\label{Sec:Z3Trotter}

As we discussed, we want to simulate the evolution due to a time-independent Hamiltonian $H$ that can be written as a sum of possibly non-commuting terms $H_j$.
\begin{equation}
	\mathfrak{U}(t) = e^{-i H t} = e^{-i (\sum_j H_j) t} = \left(e^{-i \sum_j H_j \frac{t}{M}}\right)^M.
	\label{EqSimulated}
\end{equation}	
In the last step we simply divided the total evolution into $M$ smaller time steps. In our physical system we can implement $M$ times a sequence of short evolutions involving only one of the $H_j$ terms at a time.
\begin{equation}
	\tilde{\mathfrak{U}}_M(t) = \big( \prod_j e^{-i H_j \frac{t}{M}} \big)^M.
	\label{EqSimulating}
\end{equation}	
The subscript M keeps track of the arbitrary choice of the time step. It is known that~\cite{Trotter1959a,Suzuki1985a,LLoyd1996a,Jane2003a}
\begin{equation}
	\lim_{M\rightarrow \infty} \left\| \tilde{\mathfrak{U}}_M(t) - \mathfrak{U}(t) \right\| = 0,
	\label{EqApproximation}
\end{equation}	
where $\| \cdot \|$ is the operator norm. Therefore, if all gates can be realized perfectly, the sequence $\tilde{\mathfrak{U}}_M(t)$ can approximate $\mathfrak{U}(t)$ to arbitrary precision by increasing $M$. Moreover, once we fix a particular value of $M$, we can evaluate the error due to the approximation. Bounds can be derived following reference \cite{Suzuki1985a} and adapting the proof to the case of unitary operators. We can thus get
\begin{align}
	\left\| \left(\tilde{\mathfrak{U}}_M(t) \! - \! \mathfrak{U}(t) \right)\right\| & \! \leq \! \tfrac{t^2}{2M} \big( \sum_{j<k} \left\| \left[ H_j, H_k \right] \right\| \big) {\rm exp} \big\{  \tfrac{t}{M} \Sigma_j  \left\| H_j \right\| \big\} \nonumber \\
	& \! \leq \! 3 \tfrac{t^2}{2M} \big( \sum_{j<k} \left\| \left[ H_j, H_k \right] \right\| \big)
	\label{EqSuzuki1}
\end{align}	
where the last inequality holds if $\frac{t}{M} \sum_j  \left\| H_j \right\| \leq 1$. As we discussed in Sec. II B 5, the terms $W_{ov}, W_{oh}, W_{ev}, W_{eh}$ do not commute with one another, $W_{E}$ does not commute with $W_B$, $W_{ov}, W_{oh}, W_{ev}, W_{eh}$, and $W_M$ does not commute with $W_{ov}, W_{oh}, W_{ev}, W_{eh}$. This results in     $15$ non-vanishing commutators for Eq. \eqref{EqSuzuki1}. Moreover, each commutator can be bounded by the norm of the biggest Hamiltonian piece $H_j$. We get then
\begin{align}
	\left\| \left(\tilde{\mathfrak{U}}_M(t) - \mathfrak{U}(t) \right)\right\|  \leq 45 \tfrac{t^2}{M} (\max_j \left\| H_j \right\| )^2
	\label{EqSuzuki2}
\end{align}

Now, we have that $Q$ and $P$ are unitary operators so that $\| Q \| = \| P \| = 1$.  $\| \psi^\dag(\mathbf{x}) \psi(\mathbf{x}) \| = 1$ as well, $\forall \mathbf{x}$. We define the length of the system $L$ as the number of sites along one (let's say $\hat 1$) direction. In two dimensions, the number of sites is $L^2$ and therefore $\| H_{M} \| \leq M L^2$. The number of plaquettes is $(L-1)^2$ and therefore $\| H_B \| \leq \lambda_B (L-1)^2$. The number of links is $2L(L-1)$ and therefore $\| H_E \| \leq \lambda_E L(L-1)$. The number of horizontal even links is $~(L-1)\frac{L}{2}$ and therefore $\| H_{T,h,e} \| \leq \lambda_{GM} L (L-1)$. The other gauge-matter interaction terms can be bounded in the same way. Since all the norms scale roughly as $\sim L^2$, the biggest one will be determined by the biggest coefficient $\lambda_{max}$ (some slight refinements could be needed for small size $L$). In the end, we get
\begin{align}
	\left\| \left(\tilde{\mathfrak{U}}_M(t) - \mathfrak{U}(t) \right)\right\|  \leq 45 \tfrac{L^4 t^2 \lambda_{max}^2}{M} .
	\label{EqSuzuki3}
\end{align}

\subsubsection{Number of steps and simulation time}
Our goal is to simulate the evolution of the system under $\mathfrak{U}(t)$ for a time $T$ (simulated time) and with an error (in norm) smaller than $\epsilon$. This determines the number of time slices $M$ that have to be used

\begin{equation}
	M \geq \frac{45 L^4 \lambda^2_{max} T^2}{\epsilon},
	\label{EqSteps}
\end{equation}

and the duration of each simulated time slice

\begin{equation}
	\tau_s \equiv \frac{T}{M} \leq  \frac{\epsilon}{45 L^4 \lambda^2_{max} T}.
	\label{EqSlice}
\end{equation}

Now we can compute the simulation time, i.e. the actual time taken by the experiment. We repeat $M$ times the sequence described in the previous subsection.

\begin{itemize}
	\item[1] To realize the mass Hamiltonian for the fermions, we need to give different energies to even and odd minima by shaping the optical lattice and let the system evolve for some time proportional to $\tau_s$. The actual time taken by this operation must take into account also the shaping of the lattice, that must happen adiabatically and requires a (constant) time independent of $\tau_s$ (in particular it cannot be made arbitrarily small if we want the fermions to always remain in the lowest energy level of the changing potential).
	
	\item[2] To realize the electric Hamiltonian we need to turn on and off some properly chosen laser beams and let the system evolve for some time proportional to $\tau_s$. Again, the simulation time will be linear in the simulated time, plus a constant contribution which is needed to ramp the intensity of the lasers up and down.

	\item[3] To realize the magnetic Hamiltonian we need several steps. The creation of the stators requires a constant time (that is independent of $\tau_s$). We can then evolve the auxiliary bosons for a time proportional to $\tau_s$, with an additional constant contribution which is needed to turn on and off some lasers. Finally we can undo the stators in a constant time.
	
	\item[4] To realize the gauge-matter interaction we also need several steps. First we make one of the bosonic species interact with the even fermions, for a time that is independent of of $\tau_s$. Next, we modify the fermionic potential to let them tunnel. This requires a time which is linear in $\tau_s$, apart from a constant contribution which is needed for shaping the lattice adiabatically. Finally the bosons have to interact with the fermions again.
\end{itemize}

To sum things up, the total time taken by the experiment is (for some constants $A$, $B$, $C$)

\begin{equation}
	T' = M \left( A + B \frac{T}{M}\right) = B T + C T^2
	\label{Eq:SimTime}
\end{equation}

and scales quadratically with the simulated time.

Note that the atomic collisions are typically the slowest process and will give the bigger contribution to the constant $C$. For example, in order to produce a $2\pi/3$ rotation (for the creation of stators), the required collision will take (without Feshbach resonances) about 1ms~\cite{Jane2003a}.

\subsubsection{Second-order formula}
The Trotter formula ${ e^{-i \Sigma_j H_j t} \!=\! \lim_{M \rightarrow \infty} \big( \Pi_j e^{-i H_j \frac{t}{M}} \big)^M }$ gives a very simple decomposition of the total evolution. In fact, there are infinitely many approximations of higher order~\cite{Suzuki1991a} that can be used to reduce the digitization error. For our purposes, an useful decomposition is given by the second order Trotter formula
\begin{equation}
e^{-i \Sigma_j H_j t} \!=\! \lim_{M \rightarrow \infty} \left( \prod_{j=1}^N e^{-i H_j \frac{t}{2M}} \prod_{j=N}^1 e^{-i H_j \frac{t}{2M}} \right)^M,
\end{equation}
where $N$ is the number of terms in the total Hamiltonian that can be implemented separately (8 in our specific case). This decomposition is straightforward to obtain once we know how to realize the sequence \eqref{Eq:Sequence}. First, for each of the pieces $W_j$ we have to change the time from $t/M$ to $t/2M$. Second, we must concatenate the sequence \eqref{Eq:Sequence} with its inverse. Third, we must repeat the process $M$ times. In other words, we must implement the following sequence
\begin{gather}
\left( W_{ev} W_{eh} W_{Be} W_{ov} W_{oh} W_{Bo} W_M W_E \times \right. \quad \quad \quad \quad \quad \nonumber \\
\quad \quad \quad  \times \left. W_E W_M W_{Bo} W_{oh} W_{ov} W_{Be} W_{eh} W_{ev} \right)^M,
\label{Eq:Sequence2}
\end{gather}
which requires the same experimental effort as the basic Trotter decomposition. Still, we get a much better bound for the digitization error. Indeed, if we define ${ \mathfrak{U}^{(2)}_M \equiv \big( \prod_{j=1}^N e^{-i H_j \frac{t}{2M}} \prod_{j=N}^1 e^{-i H_j \frac{t}{2M}} \big)^M }$, we have~\cite{Suzuki1985a}
\begin{align}
	\left\| \left(\tilde{\mathfrak{U}}^{(2)}_M(t) \! - \! \mathfrak{U}(t) \right)\right\| & \! \leq \! \tfrac{t^3}{M^2} \tilde{\Delta}_N \left( \{H_j\}  \right) {\rm exp} \big\{  \tfrac{t}{M} \Sigma_j  \left\| H_j \right\| \big\} \nonumber \\
	& \! \leq \! 3 \tfrac{t^3}{M^2} \tilde{\Delta}_N \left( \{H_j\}  \right),
	\label{EqSuzuki4}
\end{align}
where, following Suzuki's original notation, $\tilde{\Delta}_N \left( \{H_j\}  \right)$ is a function of nested commutators and the last inequality holds if $\frac{t}{M} \sum_j  \left\| H_j \right\| \leq 1$. Using similar arguments as before we can estimate $\tilde{\Delta}_N \left( \{H_j\}  \right)$ from above and get
\begin{align}
	\left\| \left(\tilde{\mathfrak{U}}^{(2)}_M(t) \! - \! \mathfrak{U}(t) \right)\right\| & \! \leq \! 60 \tfrac{t^3 L^6 \lambda_{max}^3}{M^2}.
	\label{EqSuzuki5}
\end{align}
Now, if we want to simulate the evolution of the system under $\mathfrak{U}(t)$ for a time $T$ (simulated time) and with an error $\epsilon$, the number of time slices $M$ that have to be used becomes
\begin{equation}
	M \geq \frac{60 L^3 \lambda^{3/2}_{max} T^{3/2}}{\sqrt{\epsilon}},
	\label{EqSteps}
\end{equation}
and the duration of each simulated time slice
\begin{equation}
	\tau_s \equiv \frac{T}{2M} \leq  \frac{\sqrt{\epsilon}}{120 L^3 \lambda^{3/2}_{max} \sqrt{T}}.
	\label{EqSlice}
\end{equation}

Note that now the number of repetition has a better scaling with respect to both the simulated time and the size of the system. Therefore, even if the number of operations required for a single time slice has doubled (compare Eq.~\eqref{Eq:Sequence2} with Eq.~\eqref{Eq:Sequence}), the total number of operations that have to be implemented becomes much smaller (for reasonable values of $L$ and $\lambda_{max} T$). This is of great help in keeping experimental errors at bay.

Moreover the simulation time, i.e. the actual time taken by the experiment is also reduced and becomes
\begin{equation}
	T' =  B T + 2 C T^{3/2},
\end{equation}
for some constants $B$, $C$ that have roughly the same value as in Eq.~\eqref{Eq:SimTime}.

\subsubsection{Higher-order bounds}
In principle, one could go even beyond the second-order formula. On one hand this allows to further reduce the number $M$ of time slices, on the other hand the number of operations required for a single time-slice increases exponentially with the order of the approximation~\cite{Berry2007a}. By balancing the two trends, it is possible to find the approximation that requires the minimum total number of operations~\cite{Berry2007a}, i.e. an experimental scheme with a shorter running time and a smaller accumulation of errors. However, formulas beyond the second-order require a much greater level of tunability of the experimental parameters.

For example, let us consider the implementation of the gauge-matter interaction, which is based on letting the fermion tunnel for a precise rescaled time $\lambda_{tun} t_{tun}$ ($\lambda_{tun}$ is here the tunnelling rate). In the case of the second order formula, we must have $\lambda_{tun} t_{tun} = T/2M$ throughout the whole experiment. Therefore, we need to implement a very specific gate and we can optimize all the relevant parameters (e.g. shaping the position and depth of neighbouring potential wells, choosing the non-rescaled tunneling time $t_{tun}$, etc.) to make this gate as precise as possible. In the case of higher order formulas, instead, the product $\lambda_{tun} t_{tun}$ has to be different for different steps of the sequence and can even take negative values~\cite{Suzuki1991a}. Therefore, we should implement a tunable gate and this would most likely give rise to bigger imprecisions in the realization. Moreover, when we want to implement steps characterized by a negative time we have actually to wait for a time $2 \pi - \lambda_{tun} t_{tun}$. This considerably increase the error done in these steps and slows down the simulation.

All things considered, using the second-order Trotter formula is a good compromise: the bounds are already much better with respect to the first-order formula and this comes practically for free.

\subsubsection{Propagation of experimental errors}
\label{Sec:Z3TrotterExp}
Each building block of the sequence \eqref{Eq:Sequence2} can in principle be affected by experimental errors of two kinds: systematic errors, that add up coherently as the same gate is repeated several times; random errors, that add up according to the central limit theorem.

Let's consider first a single gate ${ U_j(t)=e^{-i H_j t} }$. The actual realization of this gate will look like ${ U'_j(t) = e^{-i (H_j + h_j) t} }$, where  $h_j$ is the sum of a small systematic deviation and a small random fluctuation. We can assume the following condition on the norm of the fluctuations: $\| h_j \| = \delta \| H_j \| $, with $\delta \ll 1$. Then, the error (in norm) affecting the gate can be bounded as $\delta \lambda_{max} L^2 t$.

The execution of some gates (fermionic tunneling, electric Hamiltonian, etc.) takes a time proportional to $\tau_s$ and the resulting error scales as $\frac{\delta \sqrt{\epsilon}}{120 L \sqrt{\lambda_{max} T}}$ with respect to the simulation parameters.
The execution of other gates (local rotations, atomic collisions, etc.) takes a time which is independent of $\tau_s$. We can define as $t_{exp}$ a typical experimental time-scale for these gates and the resulting error scales then as $\delta \lambda_{max} L^2 t_{exp}$. This second type of error is clearly more dangerous and a lot of care has to be taken into realizing these gates in the most precise way.

We consider now the whole sequence \eqref{Eq:Sequence}. Each single gate will be repeated a number of times proportional to $2M$. Its systematic errors will add up linearly to give ${2M \frac{\delta \sqrt{\epsilon}}{120 L \sqrt{\lambda_{max} T}} = \delta L^2 \lambda_{max} T }$ or ${ 2M \delta \lambda_{max} L^2 t_{exp} = \frac{120 \; \delta \lambda_{max}^{5/2} L^5 T^{3/2} t_{exp}}{\sqrt{\epsilon}} }$ (depending on the duration of the gate). Random errors will instead follow the central limit theorem and scale with $\sqrt{2M}$ so they are less dangerous.

Finally, we can take the biggest contribution to experimental errors ($\sim 2M \delta \lambda_{max} L^2 t_{exp}$) and ask that it is comparable to the digitazion error $\epsilon$. We then find the following relation between $\delta$ and $\epsilon$
\begin{equation}
	\delta t_{exp} \sim \frac{\epsilon^{3/2}}{120 \lambda_{max}^{5/2} L^5 T^{3/2}}.
\end{equation}
This sets a strong constraint on the experimental requirements, including how they should scale with respect to the simulation "sizes" $L$ and $T$. We see that one has to reduce the timescale of the $\tau$-independent gates as much as possible, or to make the error magnitude $\delta$ as small as possible, or possibly both things simultaneously.

However, one has to keep in mind that the above bound on the errors is actually not tight, and the realization of the sequence can be much better. Indeed, it is reasonable to assume that different gates have independently distributed errors, even the systematic ones. For example, in the full sequence \eqref{Eq:Sequence} we have to realize $2M$ times the block $W_E$ and the $2M$ systematic errors will add up coherently. But then we have to add the $2M$ blocks $W_M$ and their systematic error has no correlation with the previous. Thus, there is the same probability that all $4M$ errors add up to give a big total error as that they cancel each other. By taking into account the other blocks and considering that each block is itself made of several gates, a partial cancellation of errors becomes very probable.

\subsubsection{Experimental errors}
There are several sources of error that can affect the proposed implementation. Some are common to many cold atom experiments and much effort has already been devoted to fighting them. For example, decoherence due to spontaneous scattering of lattice photons by the atoms can be slowed down to timescales of several minuites~\cite{Hamann1998,Friebel1998,Jaksch1998}.
Moving atoms around and shaping the lattice in an adiabatic way, to guarantee that atoms remain in the lowest Bloch band, has become a well-controlled technique~\cite{Jaksch1998,Aguado2008}. Other decoherence mechanisms that might result from imperfect experimental parameters, e.g. small fluctuations of the magnetic fields etc., are also commonly encountered.

 Other errors are more specific to the present proposal. For example, the digitization requires a high degree of control over the turning on/off and the duration of laser pulses, or over the overlap of atoms and their interaction time during collisions. This is crucial to obtain unitary gates that are as close as possible to the intended ones. Moreover, atomic collisions have to be elastic: the splitting of the $m_F=\pm 1$ levels has to be sufficiently different for the three atomic species, otherwise collisions will result in undesired spin flips of individual atoms.

\subsection{Possible observations}
Before moving to the concluding remarks, we would like to comment briefly on how a simple first experiment could look like. As a first step, it is easy to prepare the system in an initial state involving half-filling for the fermions (all in one sublattice as in Figure \ref{atomfig}, thus representing a Dirac sea) and with the gauge field atoms all prepared in the zero flux state ($m_F = 0$). This is the ground state of the non-interacting parts of the Hamiltonian \eqref{Eq:Hamilt}, with no static charges. Then, the interactions can be suddenly turned on (a quench), to observe and confirm the dynamical generation of only gauge invariant charge and flux configurations. This can be realised without single-site addressing of the atoms, since the atoms are initially uniformly distributed and all subsequent operations are done in parallel on the whole lattice, as detailed above.

This should be feasible with state of the art technology. Indeed, one can check that a single Trotter step corresponds to an experimental time of roughly $\sim\!25/\sim\!50 ms$, depending on the choice of the sequence \eqref{Eq:Sequence} or \eqref{Eq:Sequence2} (this time is mainly due to the creation/cancellation of stators and thus does not depend strongly on the particular time step $\tau_s$ used in the time slicing). Considering a coherence time of $\sim\!1s$ for the atomic setup, this allows to concatenate $\sim\! 20/\sim\! 40$ Trotter steps or $\sim\! 1000$ elementary gates. This is in line with other proposals of digital simulation based on ultracold atoms in optical lattices or other setups, including trapped ions~\cite{Martinez2016} and Rydberg atoms~\cite{Rydberg}. For example, in a recent experiment~\cite{Martinez2016} interesting physics was observed in a small system after an evolution involving $\sim\!200$ gate operations distributed into 4 Trotter steps.

The flux and charge configurations can be measured by fluorescence imaging~\cite{Bakr2009,Sherson2010,Torma2014}, a technique which can resolve the single occupation of each potential well in a species and $m_F$ dependent way~\cite{Boll2016}. Since different charge configurations correspond to different occupations of the fermionic wells and different flux states correspond to different $m_F$ levels of the link atoms, such measurements would reveal the complete flux and charge configuration of the whole system.

If one adds the possibility of single-site addressing of atoms, the above procedure may be repeated but, instead, starting from the ground states of the non-interacting part with static charges to observe dynamical string-breaking and perform measurements and experiments as proposed in previous proposals, such as in \cite{Zohar2015a} . A more long term goal could be to turn on the interactions in an adiabatic manner to prepare the ground state of the full model. Depending on the spectral gap of the model, this probably requires longer coherence times and further refinements of the stroboscopic sequence.

\section{Summary}

In this work, we discussed a digital formulation of lattice gauge theories, whose key is the construction of a stroboscopic, trotterized evolution of the lattice gauge theory Hamiltonian, where all the separate time steps are individually gauge invariant. This is achieved through interactions with an auxiliary system, by the use of stators.

It is possible to formulate such a digital time evolution for any lattice gauge theory based on a gauge group which is either compact Lie or finite. In the case of compact Lie gauge groups, the required local Hilbert spaces of the gauge fields have an infinite dimension, and thus they must be truncated, as usual, for a feasible quantum simulation. Conventional "representation basis" truncation schemes are not compatible with our scheme, as the stators used are based on the conjugate, ``group element basis". Thus, truncations of gauge groups should be done in group element basis, as we did in approximating $U(1)$ by $\mathbb{Z}_N$, for example.

We introduced a way to implement a $2+1$ dimensional $\mathbb{Z}_3$ lattice gauge theory using ultracold atoms in optical lattices, realizing the general ideas and concepts introduced in this
work. This complements the simulation of $\mathbb{Z}_2$, which should be even simpler, discussed in \cite{ZNsim}. The scheme proposed here allows to tailor complicated three- and four-body interactions stroboscopically from two-body interaction with an ancilla, hence reducing the need to use pertubation theory for such interaction terms \cite{AngMom}. This opens the way to possibly easier experimental realizations of a quantum simulator for lattice gauge theories in more than $1+1$ d. Another key point of the proposed scheme is the introduction of the layered structure, which allows for a clean experimental procedure: the different species may be trapped, evolved and measured separately and their interactions can be accurately controlled. Furthermore, the experimental techniques involved in our scheme (time and species dependent optical lattices, precise control of timing and overlaps during atomic collisions) are different from the experimental techniques usually required in similar proposals (as Feshbach resonances, for example). This complementary approach thus shows an alternative road toward the experimental simulation of lattice gauge theories with cold atoms, that could be less demanding, at least in some experimental setups.

\acknowledgements
The authors would like to thank Tao Shi, Andr\'as Molnar and Pranjal Bordia for helpful discussions. EZ acknowledges the support of the Alexander-von-Humboldt foundation.

%---------------APPENDIX------------------------------------------------------------
\appendix

\section{Optical lattices for $\ZZ_{3}$}
\label{App:Z3Lattices}

In this appendix, we discuss some details about the realization of the optical lattices.

\subsection{Trapping the matter fermions}
\label{App:Z3LatticesFermion}

We start with the trapping of the matter fermions, where we want to realize the following potential in the $x \minus y$ plane:

\begin{align}
	\label{Eq:Z3PotentialFermionsApp}
	V_{mat}(x,y) & \! \propto \! \frac{1}{1+f(t)+h(t)} \cos^2 \left( \pi x +\frac{\pi}{2} \right) \\
	& \plus  \frac{1}{1+g(t)+h(t)} \cos^2 \left( \pi y +\frac{\pi}{2}\right) \nonumber \\
	& \plus \frac{f(t) \plus g(t) \plus h(t)}{1\plus f(t)\plus g(t) \plus h(t)} \cos^2 \left( \frac{\pi}{2} (x \plus y) \plus \phi(t) \right). \nonumber
\end{align}

One way to implement the potential $V_{mat}(x,y)$ relies on three pairs of counter-propagating out-of-plane plain waves (lasers)

\begin{align}
	\label{Eq:Z3PotentialFermionsFields}
	\vec{E}_{mat} & \propto E_{1} \left( e^{i \vec{k}_{1} \cdot \vec{r} +i \frac{\pi}{2}} + e^{-i \vec{k}_{1} \cdot \vec{r} - i \frac{\pi}{2}} \right) \hat{e}_1 \nonumber \\
	& + E_{2} \left( e^{i \vec{k}_{2} \cdot \vec{r} +i \frac{\pi}{2}} + e^{-i \vec{k}_{2} \cdot \vec{r} - i \frac{\pi}{2}} \right) \hat{e}_2 \nonumber \\
	& + E_{3} \left( e^{i \vec{k}_{3} \cdot \vec{r} +i \phi(t)} + e^{-i \vec{k}_{3} \cdot \vec{r} - i \phi(t)} \right) \hat{e}_3 ,
\end{align}

where

\begin{gather}
	\label{Eq:Z3PotentialFermionsCoefficients}
	E_{1} = \sqrt{\tfrac{1}{1+f(t)+h(t)}}, \quad E_{2} = \sqrt{\tfrac{1}{1+g(t)+h(t)}} \nonumber \\
	E_{3}=\sqrt{\tfrac{f(t)+g(t)+h(t)}{1+f(t)+g(t)+h(t)}} \\
	\vec{k}_{1}= \pi (1,0,\xi), \quad \vec{k}_{2}= \pi (0,1,\xi), \quad \vec{k}_{3}= \pi \left(\tfrac{1}{2},\tfrac{1}{2},\zeta \right) . \nonumber
\end{gather}	

If we assume, for simplicity, that all the lasers have the same wavelength $\lambda_{mat}$ (this will depend on the particular optical transition addressed by the trap), we have to impose that ${ |\vec{k}_{1}|=|\vec{k}_{2}|=|\vec{k}_{3}|=\tfrac{2 \pi s}{\lambda_{mat}} }$ and we get the conditions ${ \zeta^{2}= \xi^{2}+\tfrac{1}{2} }$ and ${ s > \tfrac{\lambda_{mat}}{2\sqrt{2}} }$. The second condition is important in the choice of the simulated lattice spacing $s$. Next, the polarization vectors ${ \hat{e}_1, \hat{e}_2, \hat{e}_3 }$ must be orthogonal to each other. This can be achieved if

\begin{gather}
	\label{Eq:Z3PotentialFermionsPolarization}
	\hat{e}_1 \propto \{ \xi, \alpha_1, 1\}, \quad \hat{e}_2 \propto \{ \alpha_2, \xi, 1\}, \nonumber \\
	\hat{e}_3 \propto \{ \alpha_3, -\alpha_3 - 2 \sqrt{\xi^2+\tfrac{1}{2}}, 1\} ,
\end{gather}

with

\begin{gather}
	\alpha_1= \frac{1-\sqrt{1-4 \xi^4 - 2 \xi \sqrt{2 + 4 ^2}}}{2 \xi}, \nonumber \\
	\alpha_2= \frac{1+\sqrt{1-4 \xi^4 - 2 \xi \sqrt{2 + 4 \xi^2}}}{2 \xi}, \nonumber \\
	\alpha_3=\frac{-1-2\xi^2+\sqrt{1-4 \xi^4 - 2 \xi \sqrt{2 + 4 \xi^2}}}{\sqrt{2 + 4 \xi^2}}.
	\label{Eq:Z3PotentialFermionsPolarizationConditions}
\end{gather}

Note that \eq{Eq:Z3PotentialFermionsPolarizationConditions}  is only valid when ${ 1-4 \xi^4 - 2 \xi \sqrt{2 + 4 \xi^2} >0 }$. The potential generated by the fields in \eq{Eq:Z3PotentialFermionsFields} becomes

\begin{align}
	\label{Eq:Z3PotentialFermionsZ}
	V_{mat}(x,y,z) & \propto E_{1}^{2} \cos^2 \left( \pi x + \pi \xi z +\tfrac{\pi}{2} \right) \nonumber \\
	& +  E_{2}^{2}  \cos^2 \left( \pi y + \pi \xi z +\tfrac{\pi}{2}\right) \nonumber \\
	& + E_{3}^{2} \cos^2 \left( \tfrac{\pi}{2} (x+y) + \pi \zeta z +\phi(t) \right).
\end{align}

Finally, if we add a confinement along the z direction so that we can effectively put $z=0$ in \eq{Eq:Z3PotentialFermionsZ}, we get the result anticipated in \eq{Eq:Z3PotentialFermionsApp}.

\subsection{Shaping the lattice}
\label{App:Z3LatticesFermionShaping}

By tuning the functions ${ f(t), g(t), h(t), \phi(t) }$ in \eq{Eq:Z3PotentialFermionsApp} we are able to put the optical lattice in different configurations.

In the standard configuration, the lattice has equally deep minima that coincide with the sites of the simulated lattice. The minima have to be sufficiently deep to prevent tunneling and interactions between fermions residing at neighboring sites. The standard configuration can be achieved if we set ${ f(t)=g(t)=h(t)=\phi(t)=0 }$.

If we want to realize the staggering of the fermions, we need to smoothly increase the parameter $h(t)$ to some value $h_0$ (for example ${ h_0 = 0.3 }$), while keeping ${ f(t)=g(t)=\phi(t)=0 }$ as this will create an energy difference between even and odd minima. We then let the system evolve for some time and finally we bring it back smoothly to the standard configuration.

If we want to realize the tunneling of the fermions, we need to remember that this has to be done in four steps.

\begin{enumerate}
	\item For even-horizontal tunneling ($\mathcal{U}_{hop,eh}$) we need tunneling only between pairs of sites described by coordinates ${ (n, m) }$ and ${ (n+1, m) }$,
    with the additional constraint $n+m=2k$. We can smoothly increase $f(t)$ to some value $f_{0}$ (for example ${ f_0 = 2 }$)
    and tune $\phi(t)$ to the value $\pi/4$, while keeping ${ g(t)=h(t)=0 }$. In this way we lower the energy barrier between the desired sites,
     without affecting the rest of the lattice. We then let the system evolve for some time and finally we bring it back smoothly to the standard configuration.
\end{enumerate}

The remaining steps are analogous

\begin{enumerate}	
	\item[2.] For odd-horizontal tunneling (between pairs of sites described by coordinates ${ (n, m) }$ and ${ (n-1, m) }$, with  $n+m=2k$):
    Smoothly increase $f(t)$ to some value $f_{0}$ and tune $\phi(t)$ to the value $-\pi/4$, while keeping ${ g(t)=h(t)=0 }$.
	\item[3.] For even-vertical tunneling (between pairs of sites described by coordinates ${ (n, m) }$ and ${ (n, m+1) }$, with $n+m=2k$):
     Smoothly increase $g(t)$ to some value $g_{0}$ and tune $\phi(t)$ to the value $\pi/4$, while keeping ${ f(t)=h(t)=0 }$.
	\item[4.] For odd-vertical tunneling (between pairs of sites described by coordinates ${ (n, m) }$ and ${ (n, m-1) }$, with $n+m=2k$):
    Smoothly increase $g(t)$ to some value $g_{0}$ and tune $\phi(t)$ to the value $-\pi/4$, while keeping ${ f(t)=h(t)=0 }$.
\end{enumerate}	

All changes to the functions ${ f(t), g(t), h(t), \phi(t) }$ have to be implemented adiabatically~\cite{Jaksch1998,Aguado2008}, so that the atoms remain in the ground-state of the optical
 potential at all times.

\subsection{Trapping the other atoms}
\label{App:Z3LatticesOther}

Trapping the other atoms is a much easier task. For the link atoms, we need to arrange a static square lattice; for the auxiliary atoms we need a square lattice that can be moved rigidly in all three directions. This is commonly realized in current cold-atoms experiments. The only subtlety in putting the three optical lattices together is choosing the simulated lattice spacing $s$ in a way that is compatible with the condition ${ s > \tfrac{\lambda_{mat}}{2\sqrt{2}} }$ and similar conditions coming for the wavelengths ${ \lambda_{link} }$, ${ \lambda_{aux} }$ used to trap the auxiliary and the link atoms.

\begin{widetext}

\begin{figure}
  \centering
  % Requires \usepackage{graphicx}
  \includegraphics[width=0.5\textwidth]{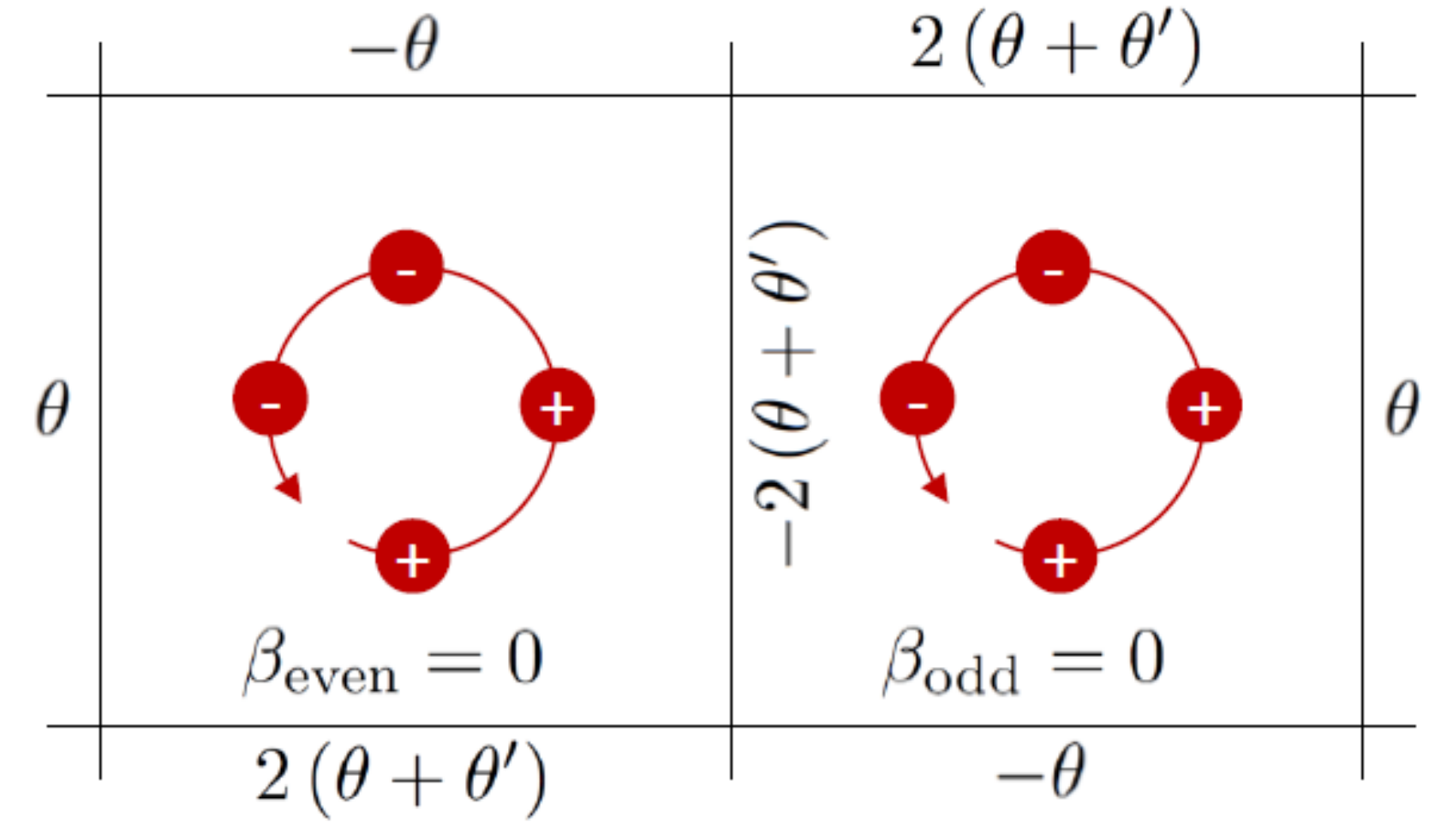}\\
  \caption{The external, static vector potential for each link, which results from the physical realization, for both an even plaquette (left) and an odd one (right). The lattice curl for both types of plaquette is zero, which means that these phases do not influence the physics and may be removed or, practically, be ignored.}\label{figphase}
\end{figure}

\section{The unphysical phases of the $\mathbb{Z}_3$ simulation}
\label{App:B}

In this appendix we wish to study the influence of the redundant phase transformations on the fermions - and show that they have no  effect whatsoever the physics -
i.e. that $W'_{oh}W'_{ov}W'_{eh}W'_{ev}$ is physically equivalent to $W_{oh}W_{ov}W_{eh}W_{ev}$. We will consider a more general case, in which the phases of the transformations
multiplying from the left and the right are different - $\theta,\theta'$; our case applies to $\theta=\theta'$. For simplicity of notation, we introduce $\epsilon = \lambda_{GM}\tau$.

We start by manipulating $W'_{ev}$,
\begin{equation}
\begin{aligned}
W'_{ev} & =e^{-i\theta'\underset{\mathbf{x}even}{\sum}\psi^{\dagger}\left(\mathbf{x}\right)\psi\left(\mathbf{x}\right)}e^{-i\epsilon\underset{\mathbf{x}even}{\sum}\left(\psi^{\dagger}\left(\mathbf{x}\right)Q\psi^{\dagger}\left(\mathbf{x}+\mathbf{\widehat{2}}\right)+h.c.\right)}e^{-i\theta\underset{\mathbf{x}even}{\sum}\psi^{\dagger}\left(\mathbf{x}\right)\psi\left(\mathbf{x}\right)}
\\ & =e^{-i\left(\theta+\theta'\right)\underset{\mathbf{x}even}{\sum}\psi^{\dagger}\left(\mathbf{x}\right)\psi\left(\mathbf{x}\right)}e^{-i\underset{\mathbf{x}even}{\sum}\left(\epsilon e^{i\theta}\psi^{\dagger}\left(\mathbf{x}\right)Q\psi^{\dagger}\left(\mathbf{x}+\mathbf{\widehat{2}}\right)+h.c.\right)}
\end{aligned}
\end{equation}
Multiplying by $W'_{eh}$, we obtain
\begin{equation}
\begin{aligned}
W'_{eh}W'_{ev}=&e^{-i\theta'\underset{\mathbf{x}even}{\sum}\psi^{\dagger}\left(\mathbf{x}\right)\psi\left(\mathbf{x}\right)}e^{-i\epsilon\underset{\mathbf{x}even}{\sum}\left(\psi^{\dagger}\left(\mathbf{x}\right)Q\psi^{\dagger}\left(\mathbf{x}+\mathbf{\widehat{1}}\right)+h.c.\right)}
e^{-i\left(2\theta+\theta'\right)\underset{\mathbf{x}even}{\sum}\psi^{\dagger}\left(\mathbf{x}\right)\psi\left(\mathbf{x}\right)}e^{-i\underset{\mathbf{x}even}{\sum}\left(\epsilon e^{i\theta}\psi^{\dagger}\left(\mathbf{x}\right)Q\psi^{\dagger}\left(\mathbf{x}+\mathbf{\widehat{2}}\right)+h.c.\right)}
\\ &=e^{-i\left(2\theta+2\theta'\right)\underset{\mathbf{x}even}{\sum}\psi^{\dagger}\left(\mathbf{x}\right)\psi\left(\mathbf{x}\right)}e^{-i\underset{\mathbf{x}even}{\sum}\left(\epsilon e^{i\left(2\theta+2\theta'\right)}\psi^{\dagger}\left(\mathbf{x}\right)Q\psi^{\dagger}\left(\mathbf{x}+\mathbf{\widehat{1}}\right)+h.c.\right)}e^{-i\underset{\mathbf{x}even}{\sum}\left(\epsilon e^{i\theta}\psi^{\dagger}\left(\mathbf{x}\right)Q\psi^{\dagger}\left(\mathbf{x}+\mathbf{\widehat{2}}\right)+h.c.\right)}
\end{aligned}
\end{equation}
Similarly,
\begin{equation}
\begin{aligned}
W'_{oh}&=e^{-i\theta'\underset{\mathbf{x}odd}{\sum}\psi^{\dagger}\left(\mathbf{x}\right)\psi\left(\mathbf{x}\right)}e^{-i\epsilon\underset{\mathbf{x}odd}{\sum}\left(\psi^{\dagger}\left(\mathbf{x}\right)Q\psi^{\dagger}\left(\mathbf{x}+\mathbf{\widehat{1}}\right)+h.c.\right)}e^{-i\theta\underset{\mathbf{x}odd}{\sum}\psi^{\dagger}\left(\mathbf{x}\right)\psi\left(\mathbf{x}\right)}
\\&=e^{-i\underset{\mathbf{x}odd}{\sum}\left(\epsilon e^{-i\theta}\psi^{\dagger}\left(\mathbf{x}\right)Q\psi^{\dagger}\left(\mathbf{x}+\mathbf{\widehat{1}}\right)+h.c.\right)}e^{-i\left(\theta+\theta'\right)\underset{\mathbf{x}odd}{\sum}\psi^{\dagger}\left(\mathbf{x}\right)\psi\left(\mathbf{x}\right)}
\end{aligned}
\end{equation}
and
\begin{equation}
\begin{aligned}
W'_{oh}W'_{ov}&=e^{-i\underset{\mathbf{x}odd}{\sum}\left(\epsilon e^{-i\theta}\psi^{\dagger}\left(\mathbf{x}\right)Q\psi^{\dagger}\left(\mathbf{x}+\mathbf{\widehat{1}}\right)+h.c.\right)}e^{-i\left(\theta+2\theta'\right)\underset{\mathbf{x}odd}{\sum}\psi^{\dagger}\left(\mathbf{x}\right)\psi\left(\mathbf{x}\right)}
e^{-i\underset{\mathbf{x}odd}{\sum}\left(\epsilon\psi^{\dagger}\left(\mathbf{x}\right)Q\psi^{\dagger}\left(\mathbf{x}+\mathbf{\widehat{2}}\right)+h.c.\right)}e^{-i\theta\underset{\mathbf{x}odd}{\sum}\psi^{\dagger}\left(\mathbf{x}\right)\psi\left(\mathbf{x}\right)}
\\&=e^{-i\underset{\mathbf{x}odd}{\sum}\left(\epsilon e^{-i\theta}\psi^{\dagger}\left(\mathbf{x}\right)Q\psi^{\dagger}\left(\mathbf{x}+\mathbf{\widehat{1}}\right)+h.c.\right)}
e^{-i\underset{\mathbf{x}odd}{\sum}\left(\epsilon e^{-i\left(2\theta+2\theta'\right)}\psi^{\dagger}\left(\mathbf{x}\right)Q\psi^{\dagger}\left(\mathbf{x}+\mathbf{\widehat{2}}\right)+h.c.\right)}e^{-i\left(2\theta+2\theta'\right)\underset{\mathbf{x}odd}{\sum}\psi^{\dagger}\left(\mathbf{x}\right)\psi\left(\mathbf{x}\right)}
\end{aligned}
\end{equation}
If we now complete the product $W'_{oh}W'_{ov}W'_{eh}W'_{ev}$, we obtain in the middle the phase
\begin{equation}
e^{-i\left(2\theta+2\theta'\right)\underset{\mathbf{x}}{\sum}\psi^{\dagger}\left(\mathbf{x}\right)\psi\left(\mathbf{x}\right)}
\end{equation}
which, within our Hilbert space (since the total number of fermions is conserved) gives rise to a global phase and thus may be ignored.

The contribution of  $W'_{oh}W'_{ov}W'_{eh}W'_{ev}$ to the sequence until is now
\begin{equation}
\begin{aligned}
e^{-i\underset{\mathbf{x}odd}{\sum}\left(\epsilon e^{-i\theta}\psi^{\dagger}\left(\mathbf{x}\right)Q\psi^{\dagger}\left(\mathbf{x}+\mathbf{\widehat{1}}\right)+h.c.\right)}e^{-i\underset{\mathbf{x}odd}{\sum}\left(\epsilon e^{-i\left(2\theta+2\theta'\right)}\psi^{\dagger}\left(\mathbf{x}\right)Q\psi^{\dagger}\left(\mathbf{x}+\mathbf{\widehat{2}}\right)+h.c.\right)}\times\\
e^{-i\underset{\mathbf{x}even}{\sum}\left(\epsilon e^{i\left(2\theta+2\theta'\right)}\psi^{\dagger}\left(\mathbf{x}\right)Q\psi^{\dagger}\left(\mathbf{x}+\mathbf{\widehat{1}}\right)+h.c.\right)}e^{-i\underset{\mathbf{x}even}{\sum}\left(\epsilon e^{i\theta}\psi^{\dagger}\left(\mathbf{x}\right)Q\psi^{\dagger}\left(\mathbf{x}+\mathbf{\widehat{2}}\right)+h.c.\right)}
\end{aligned}
\end{equation}
and we see that the gauge-matter interaction terms of $H_{GM}$ acquired some phases, which correspond
to a static external $U\left(1\right)$ vector potential $\theta\left(\mathbf{x}\right)$.
However if we calculate its lattice curl
\begin{equation}
\beta\left(\mathbf{x}\right) \equiv \theta\left(\mathbf{x},1\right) + \theta\left(\mathbf{x}+\hat{1},2\right) - \theta\left(\mathbf{x}+\hat{2},1\right) - \theta\left(\mathbf{x},2\right)
\end{equation}
which corresponds to a static magnetic field, we get that $\beta = 0$ (see Fig. \ref{figphase}).
Thus these phases have no physical effect and they can be gauged away or, effectively, ignored. We can conclude, indeed, that $W'_{oh}W'_{ov}W'_{eh}W'_{ev}$, acting on our  physical state, is equivalent to $W_{oh}W_{ov}W_{eh}W_{ev}$.

\end{widetext}
\bibliography{ref}

\end{document}